\documentclass{bmcart}

\usepackage{soul}
\usepackage{amsthm,amsmath}
\usepackage[colorlinks,linkcolor=black]{hyperref}
\usepackage{lmodern}
\usepackage{graphicx}
\usepackage[utf8]{inputenc} 
\soulregister\cite7 

\soulregister\ref7

\startlocaldefs
\endlocaldefs

\begin{document}

\begin{frontmatter}

\begin{fmbox}
\dochead{Research}


\title{A systematic evaluation of methods for cell phenotype classification using single-cell RNA sequencing data}


\author[
  addressref={aff1,aff2},                   
  email={18920927551@163.com}   
]{\inits{X.W}\fnm{Xiaowen} \snm{Cao}}
\author[
  addressref={aff3},                   
  email={lix491@usask.ca}   
]{\inits{L.X}\fnm{Li} \snm{Xing}}
\author[
  addressref={aff2,aff4},
  email={elh.majd@gmail.com}
]{\inits{E.}\fnm{Elham} \snm{Majd}}
\author[
  addressref={aff1},                   
  email={hehua@hebut.edu.cn}   
]{\inits{H.}\fnm{Hua} \snm{He}}
\author[
  addressref={aff1},                   
  email={jhgu@hebut.edu.cn}   
]{\inits{J.H}\fnm{Junhua} \snm{Gu}}
\author[
  addressref={aff2},                   
  corref={aff2},                       
  email={Xuekui@UVic.ca}   
]{\inits{X}\fnm{Xuekui} \snm{Zhang}}

\address[id=aff1]{
  \orgdiv{Department of Electrical Engineering},             
  \orgname{Hebei University of Technology},          
  \city{Tianjin},                              
  \cny{China}                                                               
}
\address[id=aff2]{
  \orgdiv{Department of Mathematics and Statistics},             
  \orgname{University of Victoria},          
    \city{Victoria},                              
  \cny{Canada}        
}

\address[id=aff4]{%
  \orgdiv{Institute of Biology},
  \orgname{National University of Sciences},
  \city{iran},
  \cny{Iran}
}
\address[id=aff3]{%
  \orgdiv{Department of Mathematics and Statistics},
  \orgname{University of Saskatchewan},
  \city{Saskatoon},
  \cny{Canada}
}



\end{fmbox}


\begin{abstractbox}

\begin{abstract} 
\parttitle{Background} 
Single-cell RNA sequencing (scRNA-seq) yields valuable insights about gene expression and gives critical information about complex tissue cellular composition. In the analysis of single-cell RNA sequencing, the annotations of cell subtypes are often done manually, which is time-consuming and irreproducible. Garnett is a cell-type annotation software based the on elastic net method. Beside cell-type annotation, supervised machine learning methods can also be applied to predict other cell phenotypes from genomic data. Despite the popularity of such applications, there is no existing study to systematically investigate the performance of those supervised algorithms in various sizes of scRNA-seq data sets. 
\parttitle{Methods and Results} 
This study evaluates 13 popular supervised machine learning algorithms to classify cell phenotypes, using published real and simulated data sets with diverse cell sizes. The benchmark contained two parts. In the first part, we used real data sets to assess the popular supervised algorithms’ computing speed and cell phenotype classification performance. The classification performances were evaluated using AUC statistics, F1-score, precision, recall, and false-positive rate. In the second part, we evaluated gene selection performance using published simulated data sets with a known list of real genes.
\parttitle{Conclusion}
The study outcomes showed that ElasticNet with interactions performed best in small and medium data sets. NB was another appropriate method for medium data sets. In large data sets, XGB works excellent. Ensemble algorithms were not significantly superior to individual machine learning methods. Adding interactions to ElasticNet can help, and the improvement was significant in small data sets.

\end{abstract}




\end{abstractbox}
%

\end{frontmatter}

\section*{Background}
In the past decade, RNA sequencing (RNA-seq) has gradually become an essential tool for analyzing differential gene expression at the full transcriptome level of mRNA splicing \cite{Svensson2018,Krzak2019}. Then, with the development of next-generation sequencing (NGS) technology, RNA-seq has been widely used \cite{Stark2019}. It has been used in many studies at the RNA level, such as single-cell gene expression, RNA translatome, and RNA structurome \cite{Slatko2018,Hu2016}. In 2013, Nature Methods named scRNA-seq as the technology of the year 2013 \cite{Editorial2014}. Then an extensive amount of scRNA-seq data was generated based on this technology. The standard analytical steps for analyzing single-cell data involve identifying the cell population in the data sets. ScRNA-seq data analysis plays a vital role in understanding the intrinsic and extrinsic cellular processes in biological and biomedical research \cite{Wang2019}. 

ScRNA-seq provides us with an opportunity to identify the cellular composition of complex tissues \cite{Svensson2018,Cheng2020}, which is valuable to detect new populations of cells, define different cell types, and discover rare cells that represent minor cell types. Since unsupervised methods often require manual annotation, which is time-consuming and cumbersome, the process of detecting new populations of cells is very tedious and inefficient \cite{Qi2020}. Supervised machine learning methods \cite{ Pliner2019} were applied to solve cell-type annotation using scRNA-seq data. ScRNA-seq classification methods predict each cell identity by learning the identities of cells from annotated training data \cite{Abdelaal2019}. Besides the method discussed above, many other supervised machine learning algorithms can be applied to classify cell phenotypes. However, there is no specific way to claim that when algorithm (A) performs better than an algorithm (B) in one specific type of problem, this superiority also should occur in other kinds of issues. Therefore, the major challenge is determining which classification algorithms have the best performance of cell-type classification using scRNA-seq data with different sample sizes.

We conducted a benchmark study to benchmark popular supervised learning algorithms to classify scRNA-seq data in various types of data with different sizes to address the mentioned challenges. 

In this study, we considered eight individual algorithms including: i) ElasticNet \cite{Zou2005}, ii) ElasticNet with interactions, iii) Linear Discriminant Analysis (LDA) \cite{Xanthopoulos2013}, iv) NaiveBayes (NB) \cite{John2013}, v) Support Vector Machines (SVM) \cite{Steinwart2008}, vi) K-Nearest Neighbors (KNN) \cite{Kramer2013}, vii) Tree \cite{Gupta2012}, and  viii) XGBoost (XGB) \cite{Friedman2003}. we constructed five ensemble algorithms based on the weights or votes of eight individual algorithms discussed above and evaluated their performance against individual ones. Ensemble learning algorithms usually work better than a specific individual algorithm \cite{Dietterich2002}. Hence, they were to apply in scRNA-seq data \cite{Abdelaal2019}. The benchmark study contained two main parts. The first part was based on classification methods. In this part, we benchmarked these method computing speed, and their performance of cell phenotype classification on three different sizes of data sets, 27 small data sets from Conquer \cite{Soneson2018}, 4 medium data sets, and 12 large ones \cite{Packer2019}. The second part was an investigation of the performance of 4 algorithms comprising i) ElasticNet, ii) NB, iii) Tree, and iv) XGB, which was used for gene selection based on the simulated data sets presented by Soneson et al. \cite{Soneson2018}. 

The comparisons were carried out from several aspects. First, we compared classical machine learning algorithms (ElasticNet with and without interaction items, LDA, NB, SVM, KNN, Tree, and XGB) with the ensemble learning algorithms (the five ensemble algorithms). Second, we investigated whether adding interaction items can optimize the results in ElasticNet. Third, we studied the performance of the algorithms in different sizes of data sets. The performance of algorithms was evaluated based on the AUC, F1-score, FPR, Precision, and Recall in processing classification. Also, we benchmarked the performance of different algorithms in gene selection by using simulated data sets. The comparison was made regarding the computation time of each algorithm.

The results showed a variation in computation time and their cell phenotype classification performance among different classification algorithms. We found that the ensemble performances in different data sets were not significantly better than the classical machine learning algorithms. ElasticNet with interactions performed well in small and medium data sets, and it spent a shorter time on classification. The results also revealed that adding interactions to the model improved the performance of ElasticNet.

Moreover, collinearity existed among genes in the data of scRNA-seq, and some genes had little significance for cell-type annotations. To make gene selection and to shrink adjustment of variable coefficients, the ElasticNet algorithm added a penalty. The penalty led to better performance among different data sets. The performance of algorithms used for gene selection in scRNA-seq data processing proved that ElasticNet was excellent in gene selection. However, when the data set sample size was large, ElasticNet needed a long time to compute. After adjusting the parameters, the results were not stable enough on some large data sets. Also, the performance of LDA and XGB were highly more effective. Overall, the results highlighted that ensemble algorithms were not superior to classical machine learning methods in speed and cell phenotype classification performance. We found ElasticNet with interactions can be more appropriate for small and medium data sets. In contrast, XGB was better for large data sets.
\section*{Methods}
This section describes data sets, algorithm design, evaluation criteria, classification methods, and gene selection.
\subsection*{Data Sets}
The data sets from Conquer \cite{Soneson2018} and GSE126954 \cite{Packer2019} were employed to evaluate all classification methods for scRNA-seq data, 27 data sets from Conquer were applied as small data sets, 4 data sets from GSE126954 were used as medium data sets, and 12 data sets from GSE126954 were considered as large ones (see Table S1, Table S2, Table S3 in Additional file 2, Additional file 3 and Additional file 4
). Finally, as the third type of data sets, the simulated data sets provided by Soneson et al. \cite{Soneson2018} were used to evaluate gene selection performance. See the details about the simulated data sets in Additional file 5 Table S4.
\subsubsection*{The 27 Data Sets from Conquer}
The conquer repository \cite{Soneson2018} was developed by Charlotte Soneson and Mark D Robinson at the University of Zurich, Switzerland. The data sets used in this research were downloaded from \url{http://imlspenticton.uzh.ch:3838/conquer}. Although Conquer contains 40 scRNA-seq data sets, we selected 27 data sets of them with two types of cells in a specific phenotype, and the number of cells in each type was at least 15. 
We selected 27 data sets in which there were two types of cells in a specific phenotype, and the number of cells in each type was at least 15 
(Please refer to Table S1 in Additional file 2
). The predictors were carefully chosen from the top 1000 genes with the strongest correlation. We named such data sets ``small data sets”.
\subsubsection*{GSE126954 Data Sets}
The GSE126954 data sets were presented by Packer et al. \cite{Packer2019}. The data sets can be downloaded from \url{https://www.ncbi.nlm.nih.gov/geo/query/acc.cgi?acc=GSE126954}. Those involved 86,024 single cells that were scRNA-seq of C.elegans embryos \cite{Packer2019}. We divided them into two parts. The first part contained 4 data sets that were larger than the small data sets from Conquer. We selected four pairs of different cells to form 4 ``medium data sets". The number of cells in our selected data sets ranged from 112 to 911 (Please refer to Table S2 in Additional file 3
for more details). The second part contained 12 data sets named ``large data sets". Besides, this paper used 2 data sets based on different cell types and 10 data sets according to a pair-wise comparison of scRNA-seq profiles of cells from C.elegans embryos at varying developmental stages. The number of each type of cell ranged from 1732 to 25875, as described in Table S3 (see Additional file 4).
. 
These two parts of the data sets were used to choose 1000 genes with the strongest correlation as predictors. Also, the algorithm parameter setting was like the setting of small data sets.
\subsubsection*{Simulated Conquer Data Sets}
To evaluate feature selection, we should know what the real gene is. In this study, we used three simulated data sets provided by Soneson et al.\cite{Soneson2018,simulated-web}. The simulation was done using the package \emph{powsimR} \cite{powsim2017}. The simulation input parameters were learned from the three real data sets: GSE45719, GSE74596, and GSE60749-GPL13112. We used information about each cell real class members received from real data sets. The simulated data sets and the real data sets can be downloaded from the website \cite{simulated-web}. Table \ref{Table5} presents more information about simulated data sets.
\subsection*{Classification Methods}
We compared the results of ElasticNet, ElasticNet with interactions, LDA, NB, SVM, KNN, Tree, XGB, and five ensemble algorithms. The first eight methods repeated 100 rounds of tenfold cross-validation. The grouping of each round of cross-validation was random. The five ensemble algorithms classification results were achieved from seven traditional algorithms, ElasticNet, LDA, NB, SVM, KNN, Tree, and XGB. We calculated the AUC, F1-score, FPR, Precision, and Recall of each algorithm performance evaluation. Also, we compared the computation time of each algorithm. Please see the results section for details.
\subsubsection*{ElasticNet}
As a combination of Ridge Regression and Lasso Regression, Elastic Network Regression can not only reduce the prediction variance but also achieve the purpose of coefficient shrinkage, and variable selection \cite{Soomro2016,Lu2021}. 
This algorithm involved two parameters:
\begin{small}
\begin{equation}
	\lambda \sum_{j=1}^{p}\left(\alpha \beta_{j}^{2}+(1-\alpha)\left|\beta_{j}\right|\right)
\end{equation}
\end{small}
To evaluate $\alpha$, we selected 6 values, $\left\{0, 0.2, 0.4, 0.6, 0.8, 1\right\}$, from 0 to 1 evenly and for evaluation of $\lambda$, we chose 100 values from $\log10^{-8}$ to $ \log10^5$. Note that the parameter of $\alpha$ and $\lambda$ in 12 large data sets were changed and set on the full data to avoid consuming a long time.
The ElasticNet model was also applied on the complete data set to select the best $\alpha$ and $\lambda$ from 12 data sets and then directly used these parameters in the tenfold cross-validation experiments. In small and medium data sets, we also considered the ElasticNet with 200 interactions. We combined 1000 genes in pairs to form interactions and then applied the logistic regression 
to find the 200 interactions with the strongest correlation and the response variable.
\subsubsection*{LDA}
Linear Discriminant Analysis (LDA) identifies a linear combination of predictors that maximize the between-class scatter and minimize the within-class scatter \cite{Park2008}. LDA uses the label information to learn a discriminant projection that can enlarge the between-class distance and reduce the within-class distance to improve the classification performance. Various extensions of LDA have also been developed to enhance the performance and efficiency \cite{Tharwat2017}.
In this benchmark study, we used the LDA function in the MASS package with parameters set to be their default values.
\subsubsection*{NB}
Naive Bayes classifier defines the probability of the document belonging to a particular class. It bases on Bayes' theorem with the assumption that features are independent. However, the Naive assumption may cause a problem because real-world features are dependent \cite{Malik2018}. We used the NaiveBayes function in the package of e1071, and the value of Laplace was set to 1.
\subsubsection*{SVM}
In SVM, which are formulated for two-class single label problems \cite{Hasan2013}. Selecting an appropriate kernel and its parameters for a specific classification problem can influence the SVM performance \cite{Hasan2017}. We adopted the ten-fold cross-validation to select the optimal parameters for $\gamma$, a parameter that comes with the function after selecting the RBF function as the kernel. It implicitly determines the distribution of the data mapped to the new feature space. The larger the $\gamma$, the fewer the support vectors, and the smaller $\gamma$ value, means the more support vectors. Moreover, the number of support vectors affects the speed of training and prediction, and also cost, it is the penalty coefficient, which is the tolerance for errors. The higher the cost, the less error can be tolerated, and it is easy to overfit. The smaller the cost, is easier for to underfit. When cost is too high or too small, the generalization ability becomes poor.

The range of $\gamma$ is $\{\frac{\left(\frac{1}{n}\right)}{10}, \left(\frac{1}{n}\right),\left(\frac{1}{n}\right) \times 10\}$, where $n$ represents the number of genes. In addition, the range of cost is $\left\{0.01, 0.1, 1, 10, 100\right\}$. Then the prediction was made with the help of the optimal model.
\subsubsection*{KNN}
This algorithm's principle is that if most of the k most similar samples to a query point q$_i$ in the feature space belong to a particular category, then a verdict can be made that the query point q$_i$ falls in this category. The distance in the feature space can measure similarity, so this algorithm is called the K-Nearest Neighbor algorithm. A train data set with accurate classification labels should be known at the beginning of the algorithm. Then for a query data q$_i$, whose label is not known and presented by a vector in the feature space, calculate the distances between it and every point in the train data set. After sorting the results of distance calculation, the class label of the test point q$_i$ can be made according to the label of the k nearest points in the train data set \cite{QKuang2009}. When the size of the train data set and test data set are both considerable, the execution time may be the bottleneck of the application \cite{QKuang2009}. 
\subsubsection*{Tree}
The Tree algorithm is a hierarchical structure that Internal Tree nodes represent splits applied to decompose the domain into regions, and terminal nodes assign class labels or class probabilities to regions believed to be sufficiently small or sufficiently uniform \cite{Kozdrowski2021}. Pruning was done by using tenfold cross-validation. Meanwhile, to avoid the situation, only one branch left after pruning would be an unpredictable classification result. The pruning was not allowed if the leaf size after pruning had been less than two.
\subsubsection*{XGB}
XGBoost is a regression Tree that has the same decision rules as a decision Tree. It supports both regression and classification. This algorithm is an efficient and scalable variant of the gradient boosting machine (GBM) \cite{Ma2020}. XGBoost method can handle sparse data, implement distributed and parallel computing flexibly \cite{Chang2018,Wang2017}.
\subsubsection*{Five ensemble algorithms}
Ensembles are achieved by generating different algorithms and combining the results into a single consensus solution\cite{Chiu2018}. This study used five ensemble algorithms constructed by the basic prediction of algorithms results applied for classification. We had two ensemble algorithms that used soft decision rule, and the other three used hard decision rule. 

We denoted $p_{n,i}$ as the predicted probability of the $n$-th sample from the $i$-th algorithm, where $i$ was the index of algorithms from \{ElasticNet, LDA, NB, SVM, KNN, Tree, XGB\}. The soft ensemble rules made a decision based on the weighted average of predicted probabilities from all methods. We named the two ensemble algorithms ensemble-weighted.AUC and ensemble-weighted.F1.

\begin{small}
\begin{equation}
\tilde{p}_{n}=\frac{\sum_{i} \text {p}_{n,i} \times w_i}
{\sum_{i} w_i}
\end{equation}
\end{small}
Where the $w_i$ represented the classification performance of each method. This paper used two criteria: F1-Median and AUC-Median as $w_i$ to construct two ensemble approaches. The ensemble method classified the $n$-th sample by the weighted probability $\tilde{p}_{n}$.

On the other hand, we denoted $O_{n,i}$ as the predicted class ($0$ or $1$) of the $n$-th sample from the $i$-th algorithm. In the hard ensemble rules, a decision was made based on the weighted votes from all methods. 
\begin{equation}
\tilde{O}_{n} =\left\{\begin{array}{l}
0; \quad \mbox{if} \quad\sum_{\{i:\text {O}_{n,i}=0\}} w_i \geq\sum_{\{i:\text {O}_{n,i}=1\}} w_i\\
1; \quad \mbox{if} \quad\sum_{\{i:\text {O}_{n,i}=0\}} w_i <\sum_{\{i:\text {O}_{n,i}=1\}} w_i
\end{array}\right.
\end{equation}
Where we chose $w_i$ as constant $1$ or AUC-Median or F1-Median, indeed, $ w_i$ was used to construct three hard ensemble rules. We named these three ensemble algorithms: ensemble-vote, ensemble-addition.AUC and ensemble-addition.F1. 
\subsection*{Design of evaluation experiments}
\subsubsection*{Cross validation}
To evaluate classification performance using supervised algorithms, we carried out tenfold cross-validation in 100 rounds after filtering genes, phenotypes, and cells. The whole samples, labeled in 0 and 1, were divided into ten groups in each round. In each of the 100 rounds of cross-validation, we applied a confusion matrix for either cell phenotype classification or gene selection. From the confusion matrixes, we calculated the following criteria to compare the performance of the methods, defined by AUC, F1-score, FPR, Precision, and Recall.

\subsubsection*{Evaluation of classification}
For each data set, one sample result in 100 rounds of experiments was predicted to be 0 or 1 (with a threshold of 0.5 by dividing the predicted results into 0 and 1). According to the prediction results and actual values of each algorithm, the confusion matrices were constructed. The numbers of samples were achieved with both real and predicted values of 0 $(a)$, actual values of 0 and predicted values of 1 $(b)$, actual values of 1 and predicted values of 0 $(c)$, and both actual and predicted values of 1 $(d)$. Recall measured the correct ratio in samples with an actual value of category 1, and the calculation formula was $d/(c+d)$ \cite{Hand2009}. The precision measured the ratio of samples with an actual value of 1 in samples, which should be 1, and the formula was $d/(b+d)$. The calculation formula of F1-score was $(2\times Recall \times Precision)/(Recall+ Precision)$. Also, the FPR measured the proportion of a sample with a prediction of 0, which was exactly 1, and finally, its calculation formula was $b/(a+b)$. AUC measured the area under the ROC curve.
\subsubsection*{Evaluation for Computation Time}
To benchmark each algorithm computation time on each data set, we recorded the computation time of each algorithm for 100 rounds and reported their average computation time in each round.

\subsubsection*{Evaluation of Gene Selection}
The applied algorithms were NB, Tree, XGB, and ElasticNet. We selected 5000 genes from the three simulated data sets provided by Soneson et al.\cite{Soneson2018,simulated-web}, which had the strongest correlation with the response variable. The system randomly selected 70$\%$ of the training sets from three simulated data sets and repeated this random selection 100 times. Then selected genes happened on 100 subsets with 5000 genes. The comparison was carried out between the selected genes and the real genes in real data sets. We will use a two-column matrix to help the calculation of each indicator. The first column is the selected genes, and the second column is the real genes. Finally, the five criteria, AUC, F1-score, FPR, Precision, and Recall, were considered to evaluate the performance of algorithms in gene selection.

Considering the Tree algorithm, we carried out cross-validation to prune the Tree with minimum deviance. Also, we applied the $Recursive Feature Elimination$\cite{Chen2020} to implement gene selection with NB. The parameter settings of the other algorithms were the same as the classification part.
\section*{Results}
In this section, we describe the detailed method comparison results using each benchmark criteria. We summarize the classification performance and the computation time of all data sets with various sample sizes. Then we summarize the performance of gene selection using simulated data sets.  
\subsection*{Classification performance}
This article studied the classification performance of different classification algorithms on 3 sample size data sets.

The results show each classification criteria to compare the performance of the methods in the rest of the subsections. This part provided three figures of the results on small data sets, medium data sets, and large data sets. Fig.\ref{fig1}\textbf{a}, Fig.\ref{fig2}\textbf{a} and Fig.\ref{fig3}\textbf{a} display one method performance, from left to right, the AUC values, F1-score values, and FPR values of the classification outcomes, as three criteria for investigating the performance of algorithms. The three criteria, AUC, F1-score and FPR values were shown in Fig.\ref{fig1}, Fig.\ref{fig2}, and Fig.\ref{fig3}. Fig.\ref{fig1}\textbf{b}, Fig.\ref{fig2}\textbf{b} and Fig.\ref{fig3}\textbf{b} represent the difference between the other algorithms and the best single algorithm. From left to right, the results display the p-value under AUC, F1-score, and FPR. The red line shows the zero value of differences among performances of algorithms. To deeply study the differences in performance of algorithms, in each figure, the discrepancies between the performance of algorithms and the best single methods were investigated using an adjusted p-value cutoff of 0.05. The p-value was considered with three decimal places. We made the values below 0.05 bold, which approved significant differences among the performance of methods.

We also proposed the values of IQR (interquartile range) and median values (the median value calculated from the results of 100 experiments for each data sets) for the five criteria in additional figures (Figure S1, Figure S2, and Figure S3 in Additional file 1). The first column shows the performance of algorithms according to five criteria, AUC, F1-score, FPR, Recall, and Precision. The second column represents the AUC-Median, F1-Median, FPR-Median, Recall-Median, and Precision-Median. The third column represents the AUC-IQR, F1-IQR, FPR-IQR, Recall-IQR, and Precision-IQR, from up to down. IQR values were used to represent the stability of the corresponding algorithm. Figure S1 illustrates the performance of 27 small data sets, Figure S2 shows the result of 4 medium data sets, and Figure S3 represents 12 large data sets (see Additional file 1).

\subsubsection*{Benchmarks for Each Classifier in Small Data Sets}
In the first part, the performance of 13 supervised algorithms was analyzed in 27 real data sets with relatively small sample sizes from Conquer study \cite{Soneson2018} used for intra-data set evaluation. The studied data sets involved relatively typical-sized scRNA-seq data sets with 24 balanced data sets and 3 unbalanced data sets (see Table S1 in Additional file 2
). The performance of 13 algorithms (ElasticNet with and without interactions, LDA, NB, SVM, KNN, Tree, XGB, and five ensemble algorithms) considering AUC, F1-score, and FPR were shown in Fig.\ref{fig1}. In Fig.\ref{fig1}\textbf{a}, each box contains 2700 scores, one score represents one value of each experiment per data set. Overall, the Fig.\ref{fig1} shows the better performance of ElasticNet with interactions as linear algorithms compared to other algorithms regarding the AUC, F1-score, and FPR. Considering the Figure S1 (see Additional file 1), the Recall of ElasticNet also was considerably better than other non-ensemble algorithms. However, the five ensemble algorithms had a higher precision than non-ensemble algorithms (see Figure S1 in Additional file 1).

As shown in Fig.\ref{fig1}\textbf{a}, the median AUC values of linear algorithms for ElasticNet with or without interactions, LDA, NB, and SVM, were over 0.987. In approximately all 27 small data sets, the median AUC values of Tree algorithms, as non-linear algorithms, were 0.967 for the Tree and 0.979 for the XGB, a higher than the KNN algorithm, with 0.966. Thus, in small data sets, the linear algorithms performed better than non-linear ones, mainly when AUC was considered the criterion.

We also used each criterion's IQR values to represent the stability of algorithms. Details were shown in Figure S1 (see Additional file 1 ). The performance of XGB was better than Tree in both performance and stability. Among the five ensemble algorithms, the two ensemble methods weighted by AUC and F1-score played a specific role in improving the results. The median AUC values for the ensemble algorithms were approximately the same, over 0.990. According to the outcomes, in small data sets, ElasticNet with interactions by the highest AUC and lower AUC-IQR was selected as the best single method. Also, this algorithm had the best performance considering  F1-score and FPR.

Fig.\ref{fig1}\textbf{b} shows the pair-wise comparison between the classification algorithms of each algorithm and the best selected single method (ElasticNet with interactions) under each criterion. There were significant differences among the performances of other algorithms and ElasticeNet with interactions as the best method that approved the preferences of ElasticNet. Moreover, considering AUC and F1-score criterion, interactions significantly improved the performance of ElasticNet without interactions. According to single methods, LDA, NB, SVM, Tree, and XGB algorithms performed significantly worse than ElasticNet with interactions.

To sum up, the investigation of the performance of 13 algorithms revealed that ElasticNet with interactions is the most practical method in small data sets. Moreover, ensemble-weighted.AUC and ensemble-weighted.F1 can be considered as the following useful methods, but not as good as ElasticNet with interactions. In comparing linear and non-linear algorithms, the linear algorithms, including ElasticNet without interactions, ElasticNet with interactions, LDA, NB, and SVM, perform better than the non-linear algorithms, and ensemble algorithms weighted by F1-score and AUC have approximately better performance among all ensemble algorithms.

\subsubsection*{Benchmarks for Each Classifier in Medium data sets}
In the second part of the classification benchmarks, we studied the 4 medium data sets from Packer et al. comprised less than 1000 samples per label (Additional file 3: Table S2
). The results helped to assess how well the supervised algorithms perform in the medium data sets. Overall, the Fig.\ref{fig2} confirmed the better performance for linear algorithms, ElasticNet with interactions and NB, rather than the non-linear algorithms and the ensemble algorithms.

As shown in Fig.\ref{fig2}\textbf{a}, the performance of ElasticNet with interactions was still better than ElasticNet without interactions considering AUC. In order to the Additional file 1: Figure S2, we also detected that the stability of ElasticNet with interactions was higher than others. But under F1-score and FPR criteria, the NB was the best single method with median values of 0.994 and 0.010, respectively. The Recall of ElasticNet and NB were also considerably better than others. Therefore, in Fig.\ref{fig2}\textbf{b}, we have marked the algorithm names under the F1-score and FPR standards in the brackets of the coordinate axis. 

According to the performance of ensemble algorithms, ensemble-weighted.AUC and ensemble-addition.F1, under AUC, F1-score, precision, and recall criteria had better performance than the other three ensemble algorithms. The results approved that, the performances of the two ensemble algorithms, ensemble-weighted.AUC and ensemble-addition.F1, had no significant preference than the best non-ensemble methods under AUC and F1-score. Thus, Fig.\ref{fig2}\textbf{a} reveals that ElasticNet with interactions performed the best under the AUC criterion, and NB had the best performance under FPR and F1-score. Moreover, ensemble-weighted.AUC and ensemble-addition.F1 were the following useful methods.

As shown in Fig.\ref{fig2}\textbf{b}, under the AUC criteria, the p-value of ensemble-weighted.F1 was 0.278, ensemble-weighted.AUC was 0.283, and ElasticNet without interactions was 0.095, which presented no significant differences between these algorithms and the AUC of ElaticNet. When F1-score and FPR were considered, NB performed slightly better. The performance of NB was significantly better than all other methods, with a p-value of almost zero. The same results were achieved under the F1-score criterion for ensemble-weighted.F1 and ensemble-weighted.AUC that approved those ensemble methods had approximately the same performances in medium data sets. 

To sum up, in medium data sets, ElasticNet with interactions seems to be more practical than others under AUC criteria. NB is the best algorithm under F1-score and FPR. The ensemble-weighted.F1 and ensemble-weighted.AUC methods can also be applied as the following suitable methods. 

\subsubsection*{Benchmarks for Each Classifier in large data sets}
In the third part of the classification benchmarks, 12 filtered data sets applied using Packer et al.’s data sets, different from the sample size of the medium data sets, generated the number of each type of cell over 1000 samples per label, ranged from 1732 to 25875, as described in Table S3 (see Additional file 1).

Fig.\ref{fig3} shows the outcomes of performance for 10 algorithms (ElasticNet, LDA, NB, Tree, XGB, and five ensemble algorithms) in large data sets, three algorithms (KNN, SVM, and ElasticNet with interactions) set aside because of long computation time. The results exposed XGB had the best performance than other single methods. Ensemble-weighted.F1 performed the best among the five ensemble algorithms under AUC, F1-score, FPR, and Recall. In terms of stability (see Additional file 1: Figure S3), the XGB, under AUC and FPR criteria, was stable but ensemble-weighted.F1 had the best stability under F1-score. Considering the Figure S3 (see Additional file 1), the Recall of XGB was also better than others. However, XGB and ensemble-weighted.F1 had the same median value of precision.

In more details, as shown in Fig.\ref{fig3}\textbf{a}, the results show that the median value of AUC for XGB was 0.999, which was higher than LDA. If the parameters of ElasticNet were set the same as medium and small data sets, it would cause the computation time to take too long, so we adjusted the parameters of ElasticNet in the large data sets. Overall, in Fig.\ref{fig3}\textbf{a}, by considering AUC, F1-score, and FPR criteria, the XGB had the best performance, while in small data sets, XGB was not a practical method. Also, the results for ElasticNet comparing to small and medium data sets dropped from 1.000 and 0.997 to 0.993 under AUC criteria, and the AUC for ElasticNet in the large data sets was at least 0.425. This situation caught our attention that described further in the section discussion. This outcome led to select XGB, in large data sets, as the best single method for comparison with the ensemble and non-ensemble methods as displayed in Fig.\ref{fig3}\textbf{b}.

Considering Fig.\ref{fig3}\textbf{b}, under the AUC and F1-score criteria, there were significant differences between the five ensemble algorithms and the best single method (XGB), the p-value was approximately zero, but, under FPR criteria, there was no significant difference among them. Thus, the other algorithms were significantly worse than the best single one (XGB), under AUC and F1 criteria, with p-values less than 0.05. ElasticNet, LDA, NB, and Tree algorithms were worse in single methods than XGB with p-values of less than 0.05.

To sum up, the investigation of the performance of 10 algorithms revealed that XGB is the most practical method for large data sets. Also, in large data sets, ensemble-weighted.F1 can also be used as an appropriate method.

\subsubsection*{Computation Time}
We organized the computation time of each algorithm on three different data sets. For details, please refer to Table \ref{Table5}. Each column represents a different algorithm, and the rows represent different data sets. The value refers to the computation time of an algorithm on a specific data set. We annotated the sample size of the data sets in Table \ref{Table5}. Underline illustrated the maximum speed.

In 27 small data sets, as shown in Table \ref{Table5}, it confirmed that SVM needed a longer time. Table \ref{Table5} displays when the sample size of the data increased to the medium-sized data sets, ElasticNet, KNN, and SVM performance slowed down. The values of computation time for the three algorithms were approximately 3.5 times, 24 times, and 10.6 times longer than the computation time of small data sets. 

By adding cross-validation to the medium data sets, SVM took a longer computation time, the average computation time was 6617.672 seconds. This result was similar to the small data sets, as shown in Table \ref{Table5}. Also, LDA and NB required short computation times, the average computation time of which were 11.570 and 21.369 seconds, respectively. When the amount of data increased to large data sets, the computation time of Tree grew little compared to KNN.

Considering each algorithm computation time, KNN occupied too much memory in the computation time of large data sets. Therefore, we could not commonly compare these time-consuming and labor-intensive algorithms, ElasticNet, KNN, and SVM. Hence, we adjusted the parameters of ElasticNet for large data sets. In the large data set, we calculated the computation time of 5 algorithms: the adjusted ElasticNet, LDA, NB, Tree, and XGB (see Table \ref{Table5}). The average computation time of XGB, as the best single method in large data sets, was 2803.485 seconds.
\subsection*{Gene Selection}
The dimensions of gene variables in the scRNA-seq data sets were large. When we implemented the classification prediction, some algorithms automatically made gene selection for data sets. Hence, we studied whether there was a particular relationship between gene selection performance and classification prediction performance. Therefore, we used four algorithms, NB, Tree, XGB, and ElasticNet, suitable for gene selection, using the three simulated data sets provided by Soneson et al. \cite{Soneson2018} (Please see section methods for more details).

We set different seeds in each round for each data set and performed gene selections 100 times on 100 subsets with 70$\%$ of cells. Then, the results were compared with the genes in the real data sets, GSE45719, GSE74596, and GSE60749-GPL13112 (Additional file 5: Table S4
), and the final evaluation results were obtained.

Fig.\ref{fig4} displays the AUC values of the four algorithms, NB, Tree, XGB, and ElasticNet, for gene selection in all experiments of three data sets. We marked the name of the data sets above Fig.\ref{fig4}. In each box of Fig.\ref{fig4}, 100 points show the 100 subsets. Considering the AUC criteria, as shown in Fig.\ref{fig4}, the median value of AUC for the four algorithms on the simulated data set of GSE60749-GPL13112 was approximately 0.500. In the other two data sets, the median value of the AUC for ElasticNet was about 0.711 and 0.637, which revealed better gene selection than those three algorithms.
\section*{Discussion}
This study evaluated and compared the classification performance in various data sets of different sizes and each algorithm's computation time. 27 Conquer real data sets proposed by Soneson et al. \cite{Soneson2018} were used for classification as small data sets, as shown in Table S1 (see Additional file 2)
, and 4 data sets presented by Packer et al. \cite{Packer2019}, as shown in Table S2 (see Additional file 3)
which were considered medium data sets. The 12 large data sets presented by Packer et al. \cite{Packer2019} were also applied for classification, as shown in Table S3 (see Additional file 4)
. We also investigated gene selection. Three simulated data sets presented by Soneson et al. \cite{Soneson2018} were used for gene selection, as shown in Table S4 (see Additional file 5).

In this paper, we conducted some cross-validation experiments and gene selection, provided detailed information on data sets, compared the performance of various algorithms in many aspects such as performance, stability, computation time, and gene selection, and finally summarized each algorithm's performance on different sizes of data sets. According to outcomes, ElasticNet with interactions was more suitable for processing small and medium scRNA-seq data. However, considering the AUC criterion, interactions improved ElasticNet in small and medium data sets. There were significant differences between ElasticNet with and without interactions in small data sets. When ElasticNet was used in large data sets with and without interactions, they acted slightly inappropriate because of the non-convergence issue. Hence, we adjusted the parameters, but they had a long computation time. Therefore, we only studied ElasticNet and ignored ElasticNet with interactions. Moreover, NB can also be selected as the practical algorithm in medium data sets if F1-score and FPR be considered. The results in large data sets approved that if the computation time is ignored, then XGB can be regarded as the best algorithm. Moreover, the study illustrated that the ensemble algorithms were not always better than the classic linear ones. We also found that the integrated methods did not significantly improve the performance of the single algorithms.
\subsection*{Classification on small data sets}
In small data sets, the performance of 13 algorithms was evaluated and compared using five criteria, AUC, F1-score, FPR, Precision, and Recall. Considering AUC as the standard (see Fig.\ref{fig1}). We observed that ElasticNet, ElasticNet with interactions, SVM, and LDA performed well, but LDA was unstable. The computation time of the SVM was too long. ElasticNet with interactions was a little more suitable for classification than ElasticNet without interactions in terms of performance and stability. The results showed a higher median value of AUC and a lower value of AUC-IQR for ElasticNet with interactions. Thus, ElasticNet with interactions can be considered as the best linear algorithm for small data sets. XGB performance was better than Tree,considering to the value of AUC and AUC-IQR. The ensemble algorithms can also improve the overall generalization ability when each base learner's performance was more significant than 0.5. Therefore, it is understandable that some kinds of ensemble strategies perform better than others. Although, we found no significant better performance for ensemble methods compared to single methods. 

SVM performed poorly when F1-score and FPR were considered as the standard. It seems that this was because of calculating both performance and stability in those criteria. Also, SVM performed very slowly. It may cause that the algorithm involves the selection of several optimal parameters, so this result is logical. Both LDA and NB were simple linear classifiers with fast computation time, and their classification performances were good. ElasticNet and ElasticNet with interactions took longer than LDA, and the median value of computation time was even seven times more than NB but eleven times less than SVM. Like SVM, ElasticNet and ElasticNet with interactions were also needed to select the optimal parameters to decrease the computation time. But such computation time can be accommodated on small data sets. Therefore, we suggest applying ElasticNet with interactions with high prediction performance and stability for small data sets.

\subsection*{Classification on medium data sets}
On medium data sets, we compared the performance of 13 algorithms. According to AUC as the standard, ElasticNet with interactions performed well with higher performance and stability, similar to the small data sets. LDA and SVM did not perform well on these four data sets. The reason can be to focus on only 4 data sets, which may happen by chance. XGB was also better than Tree in both performance and stability. Also, two ensemble algorithms, ensemble-weighted.AUC and ensemble-weighted.F1, performed better. However, the two ensemble algorithms were not significantly different from the single algorithm (ElasticNet with interactions) under AUC. Considering the median value of AUC, KNN still performed poorly than others. We found KNN is not suitable for processing scRNA-seq data because of taking up more memory. As the number of samples increased, the amount of calculation of KNN also rose. Hence, insufficient memory during the calculation process happened.

Under the F1-score and FPR standard, the performance of NB was more prominent, better than other algorithms. The performances of other algorithms were similar as under the AUC criterion. Overall, according to each criterion, the linear algorithms continued to overcome the non-linear algorithms in medium data sets.

Therefore, we suggest that ElasticNet with interactions and NB are better to apply in medium data sets because of better performance and stability under each criterion and accommodate consumption time. 

\subsection*{Classification on large data sets}
In large data sets, the performance of 10 algorithms was compared (see Fig.\ref{fig3}). Using AUC as the standard, the performance and computation time of ElasticNet, LDA, and SVM took longer than others. Thus, SVM was removed, and the parameters in ElasticNet were modified. We found XGB is the best single algorithm in large data sets. ElasticNet with interactions was not considered in the final results because of its long computation time. However, we found the performance was not suitable due to some non-convergence issues. It is worth noting that ElasticNet still performed well in the convergence experiments. Therefore, ElasticNet may also be the right choice after solving the non-convergence problem. Two ensemble algorithms performed the best as they did in small data sets. The core idea of the KNN algorithm is that if most of the K-Nearest samples in the feature space of a sample belong to a particular category, the sample also belongs to that category and has the samples' characteristics. Hence, when the size of data sets increases, more computation is needed \cite{Salvador-Meneses2019}. In this paper, KNN was not considered due to out-of-memory issues. 

Overall, according to three criteria, AUC, F1-score, and FPR, the performance of XGB was almost prominent, even more than other algorithms. Thus, we suggest XGB for large data sets because of better performance and stability.

From a computation time point of view, KNN and SVM took a long computation time. Therefore, they were not under our consideration. The computation time problem of ElasticNet has been improved compared to when the parameters were not adjusted, but it was still long (please check Table \ref{Table5}).
Hence, the results illustrated that the non-convergence issue leads to more extended computation. Moreover, the computation time of XGB is higher than linear classifiers (LDA and NB) because there is a complicated calculation in XGB.

This article investigated whether there is a correlation between ``perfect separation" and ``model non-convergence." In this case, we collected statistics on convergence and perfect separation in each round of 1,000 experiments on 12 large data sets. The result revealed that over 69$\%$ of the experiments showed ``non-convergence and perfect separation" or ``convergence without perfect separation." Therefore, we claim that there is a relationship between 
``perfect separation" and ``model non-convergence." Moreover, combined with the computation time and performance, we claim LDA and XGB are more suitable for large data sets because of faster computation time and better performance.

Finally, we discovered that the linear algorithms were better than the non-linear ones to classify different-sized data sets. Among all, ElasticNet with interactions performed well on small and medium data sets. However, NB also was practical in medium data sets, and XGB worked better in large data sets. Also, there was no preference in the performance of ensemble and single methods.
\subsection*{Gene selection on simulated data sets}
In this study, the 100 subsets in three simulated data sets, provided by Soneson et al.\cite{Soneson2018}, were used to avoid a certain level of randomness. Those 100 random subsets were received by 100 times of experiments. The selected genes were compared with the real genes of real data sets. The results revealed ElasticNet performed the best (please  
see Fig.\ref{fig4}). 
ElasticNet achieves feature selection by regulating non-relevant predictors' coefficients to zero \cite{Xuekui2020} and enhanced the gene selections on scRNA-seq data. Thus, it was reasonable that it performed the best.
\section*{Conclusions}
In this paper, a comprehensive evaluation of supervised algorithms for the classification of scRNA-seq data was presented to assess the performance of those algorithms in different sizes of data sets. Several algorithms were used, including i) ElasticNet, ii) ElasticNet with interactions, iii) LDA, iv) NB, v) SVM, vi) KNN, vii) Tree, viii) XGB and x) five ensemble algorithms based on the weights and votes given to the seven algorithms, ElasticNet, LDA, NB, SVM, KNN, Tree, and XGB. Enormous differences were determined in the performance of algorithms in response to changing the input features. The computation times of algorithms were considerably varied based on the number of cells, features, and the type of algorithms.

According to the outcomes, ElasticNet with interactions had better performance for small and medium data sets, while XGB was suitable for large data sets. The gene selection performance of ElasticNet was perfect when comparing the genes selected by the four algorithms on the three simulated data sets with the genes in the actual data sets. The outcomes revealed that ensemble algorithms were not always superior to classical machine learning methods in performance and computation time. Also, the results approved that there were significant differences between ElasticNet with interactions than without interactions.

We recommend ElasticNet with supervised learning classification interactions in small and medium data sets because it performs better than the other classifiers. Interactions improved the performance of ElasticNet significantly in small data sets. NB can be considered as a proper classifier for medium data sets as well. However, when the sample size of the data sets was large, XGB was more suitable. Although the computation time of XGB was slightly longer, the performance was relatively higher than other methods. There was no reason to encourage using ensemble algorithms while there was no application priority other than single methods.

\begin{backmatter}

\section*{Acknowledgements}
This research was enabled in part by support provided by WestGrid (www.westgrid.ca) and Compute Canada (www.computecanada.ca).

\section*{Funding}
This work was supported by the Natural Sciences and Engineering Research Council Discovery Grants (RGPIN‐2017‐04722 PI:XZ), the Canada Research Chair (950‐231363, PI:XZ). 

\section*{Abbreviations}
Not applicable

\section*{Availability of data and materials}
27 small data sets used in this article can be obtained from \url{https://github.com/markrobinsonuzh/conquer}, and the medium and large data sets from GSE126954 can be obtained from \url{https://www.ncbi.nlm.nih .gov/geo/query/acc.cgi?acc=GSE126954}, the simulation data sets based on GSE45719, GSE74596, and GSE60749-GPL13112 can be obtained from \url{http://imlspenticton.uzh.ch/robinson_lab/conquer_de_comparison/}

\section*{Ethics approval and consent to participate}
Not applicable.

\section*{Competing interests}
The authors declare that they have no competing interests.

\section*{Consent for publication}
Not applicable.

\section*{Author's contributions}
Li Xing and Xuekui Zhang contributed to the study concept and design. Xiaowen Cao contributed to data processing, data analysis, and preparation of the first draft. All authors have developed drafts of the manuscript, approved the final draft of the manuscript. Xuekui Zhang supervised this project.


\bibliographystyle{bmc-mathphys} 
\bibliography{bmc_article}      


\begin{thebibliography}{40}
\ifx \bisbn   \undefined \def \bisbn  #1{ISBN #1}\fi
\ifx \binits  \undefined \def \binits#1{#1}\fi
\ifx \bauthor  \undefined \def \bauthor#1{#1}\fi
\ifx \batitle  \undefined \def \batitle#1{#1}\fi
\ifx \bjtitle  \undefined \def \bjtitle#1{#1}\fi
\ifx \bvolume  \undefined \def \bvolume#1{\textbf{#1}}\fi
\ifx \byear  \undefined \def \byear#1{#1}\fi
\ifx \bissue  \undefined \def \bissue#1{#1}\fi
\ifx \bfpage  \undefined \def \bfpage#1{#1}\fi
\ifx \blpage  \undefined \def \blpage #1{#1}\fi
\ifx \burl  \undefined \def \burl#1{\textsf{#1}}\fi
\ifx \doiurl  \undefined \def \doiurl#1{\textsf{#1}}\fi
\ifx \betal  \undefined \def \betal{\textit{et al.}}\fi
\ifx \binstitute  \undefined \def \binstitute#1{#1}\fi
\ifx \binstitutionaled  \undefined \def \binstitutionaled#1{#1}\fi
\ifx \bctitle  \undefined \def \bctitle#1{#1}\fi
\ifx \beditor  \undefined \def \beditor#1{#1}\fi
\ifx \bpublisher  \undefined \def \bpublisher#1{#1}\fi
\ifx \bbtitle  \undefined \def \bbtitle#1{#1}\fi
\ifx \bedition  \undefined \def \bedition#1{#1}\fi
\ifx \bseriesno  \undefined \def \bseriesno#1{#1}\fi
\ifx \blocation  \undefined \def \blocation#1{#1}\fi
\ifx \bsertitle  \undefined \def \bsertitle#1{#1}\fi
\ifx \bsnm \undefined \def \bsnm#1{#1}\fi
\ifx \bsuffix \undefined \def \bsuffix#1{#1}\fi
\ifx \bparticle \undefined \def \bparticle#1{#1}\fi
\ifx \barticle \undefined \def \barticle#1{#1}\fi
\ifx \bconfdate \undefined \def \bconfdate #1{#1}\fi
\ifx \botherref \undefined \def \botherref #1{#1}\fi
\ifx \url \undefined \def \url#1{\textsf{#1}}\fi
\ifx \bchapter \undefined \def \bchapter#1{#1}\fi
\ifx \bbook \undefined \def \bbook#1{#1}\fi
\ifx \bcomment \undefined \def \bcomment#1{#1}\fi
\ifx \oauthor \undefined \def \oauthor#1{#1}\fi
\ifx \citeauthoryear \undefined \def \citeauthoryear#1{#1}\fi
\ifx \endbibitem  \undefined \def \endbibitem {}\fi
\ifx \bconflocation  \undefined \def \bconflocation#1{#1}\fi
\ifx \arxivurl  \undefined \def \arxivurl#1{\textsf{#1}}\fi
\csname PreBibitemsHook\endcsname

\bibitem{Svensson2018}
\begin{barticle}
\bauthor{\bsnm{Svensson}, \binits{V.}},
\bauthor{\bsnm{Vento-Tormo}, \binits{R.}},
\bauthor{\bsnm{Teichmann}, \binits{S.A.}}:
\batitle{Exponential scaling of single-cell rna-seq in the past decade}.
\bjtitle{Nature Protocols}
\bvolume{13},
\bfpage{599}--\blpage{604}
(\byear{2018})
\end{barticle}
\endbibitem

\bibitem{Krzak2019}
\begin{barticle}
\bauthor{\bsnm{Krzak}, \binits{M.}},
\bauthor{\bsnm{Raykov}, \binits{Y.}},
\bauthor{\bsnm{Boukouvalas}, \binits{A.}},
\bauthor{\bsnm{Cutillo}, \binits{L.}},
\bauthor{\bsnm{Claudia}, \binits{A.}}:
\batitle{Benchmark and parameter sensitivity analysis of single-cell rna
  sequencing clustering methods}.
\bjtitle{Frontiers in Genetics}
\bvolume{10},
\bfpage{1}--\blpage{19}
(\byear{2019})
\end{barticle}
\endbibitem

\bibitem{Stark2019}
\begin{barticle}
\bauthor{\bsnm{Stark}, \binits{R.}},
\bauthor{\bsnm{Grzelak}, \binits{M.}},
\bauthor{\bsnm{Hadfield}, \binits{J.}}:
\batitle{Rna sequencing: the teenage years}.
\bjtitle{Nature Reviews Genetics}
\bvolume{20},
\bfpage{631}--\blpage{656}
(\byear{2019})
\end{barticle}
\endbibitem

\bibitem{Slatko2018}
\begin{barticle}
\bauthor{\bsnm{Slatko}, \binits{B.E.}},
\bauthor{\bsnm{Gardner}, \binits{A.F.}},
\bauthor{\bsnm{Ausubel}, \binits{F.M.}}:
\batitle{Overview of next generation sequencing technologies}.
\bjtitle{Molecular Biology}
\bvolume{122},
\bfpage{1}--\blpage{15}
(\byear{2018})
\end{barticle}
\endbibitem

\bibitem{Hu2016}
\begin{barticle}
\bauthor{\bsnm{Hu}, \binits{Y.}},
\bauthor{\bsnm{Hase}, \binits{T.}},
\bauthor{\bsnm{Li}, \binits{H.P.}},
\bauthor{\bsnm{Prabhakar}, \binits{S.}},
\bauthor{\bsnm{Kitano}, \binits{H.}},
\bauthor{\bsnm{Ng}, \binits{S.K.}},
\bauthor{\bsnm{Ghosh}, \binits{S.}},
\bauthor{\bsnm{Wee}, \binits{L.J.K.}}:
\batitle{A machine learning approach for the identification of key markers
  involved in brain development from single-cell transcriptomic data}.
\bjtitle{BMC genomics}
\bvolume{17},
\bfpage{1025}
(\byear{2016})
\end{barticle}
\endbibitem

\bibitem{Editorial2014}
\begin{botherref}
{Method of the Year 2013}.
Nature Methods
\textbf{11}(1),
1
(2014).
doi:\doiurl{10.1038/nmeth.2801}
\end{botherref}
\endbibitem

\bibitem{Wang2019}
\begin{barticle}
\bauthor{\bsnm{Wang}, \binits{T.}},
\bauthor{\bsnm{Li}, \binits{B.}},
\bauthor{\bsnm{Nelson}, \binits{C.E.}},
\bauthor{\bsnm{Nabavi}, \binits{S.}}:
\batitle{Comparative analysis of differential gene expression analysis tools
  for single-cell rna sequencing data}.
\bjtitle{BMC bioinformatics}
\bvolume{20}(\bissue{4}),
\bfpage{1}--\blpage{16}
(\byear{2019})
\end{barticle}
\endbibitem

\bibitem{Cheng2020}
\begin{barticle}
\bauthor{\bsnm{Cheng}, \binits{Q.}},
\bauthor{\bsnm{Li}, \binits{J.}},
\bauthor{\bsnm{Fan}, \binits{F.}},
\bauthor{\bsnm{Cao}, \binits{H.}},
\bauthor{\bsnm{Dai}, \binits{Z.Y.}},
\bauthor{\bsnm{Wang}, \binits{Z.Y.}},
\bauthor{\bsnm{Feng}, \binits{S.S.}}:
\batitle{Identification and analysis of glioblastoma biomarkers based on single
  cell sequencing}.
\bjtitle{Frontiers in Bioengineering and Biotechnology}
\bvolume{8},
\bfpage{1}--\blpage{7}
(\byear{2020})
\end{barticle}
\endbibitem

\bibitem{Qi2020}
\begin{barticle}
\bauthor{\bsnm{Qi}, \binits{R.}},
\bauthor{\bsnm{Ma}, \binits{A.}},
\bauthor{\bsnm{Ma}, \binits{Q.}},
\bauthor{\bsnm{Zou}, \binits{Q.}}:
\batitle{Clustering and classification methods for single-cell rna-sequencing
  data}.
\bjtitle{Briefings in bioinformatics}
\bvolume{21},
\bfpage{1196}--\blpage{1208}
(\byear{2020})
\end{barticle}
\endbibitem

\bibitem{Pliner2019}
\begin{barticle}
\bauthor{\bsnm{Pliner}, \binits{H.A.}},
\bauthor{\bsnm{Shendure}, \binits{J.}},
\bauthor{\bsnm{Trapnell}, \binits{C.}}:
\batitle{Supervised classification enables rapid annotation of cell atlases}.
\bjtitle{Nature Methods}
\bvolume{16},
\bfpage{983}--\blpage{986}
(\byear{2019})
\end{barticle}
\endbibitem

\bibitem{Abdelaal2019}
\begin{barticle}
\bauthor{\bsnm{Abdelaal}, \binits{T.}},
\bauthor{\bsnm{Michielsen}, \binits{L.}},
\bauthor{\bsnm{Cats}, \binits{D.}},
\bauthor{\bsnm{Hoogduin}, \binits{D.}},
\bauthor{\bsnm{Mei}, \binits{H.}},
\bauthor{\bsnm{Reinders}, \binits{M.}},
\bauthor{\bsnm{Mahfouz}, \binits{A.}}:
\batitle{A comparison of automatic cell identification methods for single-cell
  rna-sequencing data}.
\bjtitle{Genome Biology}
\bvolume{20},
\bfpage{194}
(\byear{2019})
\end{barticle}
\endbibitem

\bibitem{Zou2005}
\begin{barticle}
\bauthor{\bsnm{Zou}, \binits{H.}},
\bauthor{\bsnm{Hastie}, \binits{T.}}:
\batitle{Regularization and variable selection via the elastic net}.
\bjtitle{Journal of the royal statistical society: series B (statistical
  methodology)}
\bvolume{67},
\bfpage{301}--\blpage{320}
(\byear{2005})
\end{barticle}
\endbibitem

\bibitem{Xanthopoulos2013}
\begin{bchapter}
\bauthor{\bsnm{Xanthopoulos}, \binits{P.}},
\bauthor{\bsnm{Pardalos}, \binits{P.M.}},
\bauthor{\bsnm{Trafalis}, \binits{T.B.}}:
\bctitle{Linear discriminant analysis}.
In: \bbtitle{Robust Data Mining},
pp. \bfpage{27}--\blpage{33}.
\bpublisher{Springer},
\blocation{USA, New York}
(\byear{2013})
\end{bchapter}
\endbibitem

\bibitem{John2013}
\begin{botherref}
\oauthor{\bsnm{John}, \binits{G.H.}},
\oauthor{\bsnm{Langley}, \binits{P.}}:
Estimating continuous distributions in bayesian classifiers.
arXiv preprint arXiv:1302.4964,
338--345
(2013)
\end{botherref}
\endbibitem

\bibitem{Steinwart2008}
\begin{bbook}
\bauthor{\bsnm{Steinwart}, \binits{I.}},
\bauthor{\bsnm{Christmann}, \binits{A.}}:
\bbtitle{Support Vector Machines}.
\bpublisher{Springer},
\blocation{Germany}
(\byear{2008})
\end{bbook}
\endbibitem

\bibitem{Kramer2013}
\begin{bchapter}
\bauthor{\bsnm{Kramer}, \binits{O.}}:
\bctitle{K-nearest neighbors}.
In: \bbtitle{Dimensionality Reduction with Unsupervised Nearest Neighbors},
pp. \bfpage{13}--\blpage{23}.
\bpublisher{Springer},
\blocation{USA, New York}
(\byear{2013})
\end{bchapter}
\endbibitem

\bibitem{Gupta2012}
\begin{barticle}
\bauthor{\bsnm{Gupta}, \binits{D.}},
\bauthor{\bsnm{Malviya}, \binits{A.}},
\bauthor{\bsnm{Singh}, \binits{S.}}:
\batitle{Performance analysis of classification tree learning algorithms}.
\bjtitle{International Journal of Computer Applications}
\bvolume{55}(\bissue{6}),
\bfpage{39}--\blpage{44}
(\byear{2012})
\end{barticle}
\endbibitem

\bibitem{Friedman2003}
\begin{barticle}
\bauthor{\bsnm{Friedman}, \binits{J.H.}},
\bauthor{\bsnm{Meulman}, \binits{J.J.}}:
\batitle{Multiple additive regression trees with application in epidemiology}.
\bjtitle{Statistics in medicine}
\bvolume{22},
\bfpage{1365}--\blpage{1381}
(\byear{2003})
\end{barticle}
\endbibitem

\bibitem{Dietterich2002}
\begin{barticle}
\bauthor{\bsnm{Dietterich}, \binits{T.G.}}:
\batitle{Ensemble learning}.
\bjtitle{The handbook of brain theory and neural networks}
\bvolume{2},
\bfpage{110}--\blpage{125}
(\byear{2002})
\end{barticle}
\endbibitem

\bibitem{Soneson2018}
\begin{barticle}
\bauthor{\bsnm{Soneson}, \binits{C.}},
\bauthor{\bsnm{Robinson}, \binits{M.D.}}:
\batitle{Bias, robustness and scalability in single-cell differential
  expression analysis}.
\bjtitle{Nature Methods}
\bvolume{15},
\bfpage{255}--\blpage{261}
(\byear{2018})
\end{barticle}
\endbibitem

\bibitem{Packer2019}
\begin{barticle}
\bauthor{\bsnm{Packer}, \binits{J.S.}},
\bauthor{\bsnm{Zhu}, \binits{Q.}},
\bauthor{\bsnm{Huynh}, \binits{C.}},
\bauthor{\bsnm{Sivaramakrishnan}, \binits{P.}},
\bauthor{\bsnm{Preston}, \binits{E.}},
\bauthor{\bsnm{Dueck}, \binits{H.}},
\bauthor{\bsnm{Stefanik}, \binits{D.}},
\bauthor{\bsnm{Tan}, \binits{K.}},
\bauthor{\bsnm{Trapnell}, \binits{C.}},
\bauthor{\bsnm{Kim}, \binits{J.}},
\bauthor{\bsnm{Waterston}, \binits{R.H.}},
\bauthor{\bsnm{Murray}, \binits{J.I.}}:
\batitle{A lineage-resolved molecular atlas of c. elegans embryogenesis at
  single-cell resolution}.
\bjtitle{Science}
\bvolume{365},
\bfpage{1}--\blpage{15}
(\byear{2019})
\end{barticle}
\endbibitem

\bibitem{simulated-web}
\begin{botherref}
\oauthor{\bsnm{Soneson}, \binits{C.}},
\oauthor{\bsnm{Robinson}, \binits{M.D.}}
\url{http://imlspenticton.uzh.ch/robinson_lab/conquer_de_comparison}
\end{botherref}
\endbibitem

\bibitem{powsim2017}
\begin{barticle}
\bauthor{\bsnm{Vieth}, \binits{B.}},
\bauthor{\bsnm{Ziegenhain}, \binits{C.}},
\bauthor{\bsnm{Parekh}, \binits{S.}},
\bauthor{\bsnm{Enard}, \binits{W.}},
\bauthor{\bsnm{Hellmann}, \binits{I.}}:
\batitle{Powsimr: Power analysis for bulk and single cell rna-seq experiments}.
\bjtitle{Bioinformatics}
\bvolume{33}(\bissue{21}),
\bfpage{3486}--\blpage{3488}
(\byear{2017})
\end{barticle}
\endbibitem

\bibitem{Soomro2016}
\begin{barticle}
\bauthor{\bsnm{Soomro}, \binits{B.N.}},
\bauthor{\bsnm{Xiao}, \binits{L.}},
\bauthor{\bsnm{Huang}, \binits{L.}},
\bauthor{\bsnm{Soomro}, \binits{S.H.}},
\bauthor{\bsnm{Molaei}, \binits{M.}}:
\batitle{Bilayer elastic net regression model for supervised spectral-spatial
  hyperspectral image classification}.
\bjtitle{IEEE Journal of Selected Topics in Applied Earth Observations and
  Remote Sensing}
\bvolume{9}(\bissue{9}),
\bfpage{4102}--\blpage{4116}
(\byear{2016})
\end{barticle}
\endbibitem

\bibitem{Lu2021}
\begin{barticle}
\bauthor{\bsnm{Lu}, \binits{J.}},
\bauthor{\bsnm{Chen}, \binits{M.}},
\bauthor{\bsnm{Qin}, \binits{Y.}}:
\batitle{Drug-induced cell viability prediction from lincs-l1000 through
  wrfen-xgboost algorithm}.
\bjtitle{BMC bioinformatics}
\bvolume{22}(\bissue{1}),
\bfpage{1}--\blpage{18}
(\byear{2021})
\end{barticle}
\endbibitem

\bibitem{Park2008}
\begin{barticle}
\bauthor{\bsnm{Park}, \binits{C.H.}},
\bauthor{\bsnm{Park}, \binits{H.}}:
\batitle{A comparison of generalized linear discriminant analysis algorithms}.
\bjtitle{Pattern Recognition}
\bvolume{41}(\bissue{3}),
\bfpage{1083}--\blpage{1097}
(\byear{2008})
\end{barticle}
\endbibitem

\bibitem{Tharwat2017}
\begin{barticle}
\bauthor{\bsnm{Tharwat}, \binits{A.}},
\bauthor{\bsnm{Gaber}, \binits{T.}},
\bauthor{\bsnm{Ibrahi}, \binits{A.}},
\bauthor{\bsnm{Hassanien}, \binits{A.E.}}:
\batitle{Linear discriminant analysis: A detailed tutorial}.
\bjtitle{AI communications}
\bvolume{30}(\bissue{2}),
\bfpage{169}--\blpage{190}
(\byear{2017})
\end{barticle}
\endbibitem

\bibitem{Malik2018}
\begin{barticle}
\bauthor{\bsnm{Malik}, \binits{V.}},
\bauthor{\bsnm{Kumar}, \binits{A.}}:
\batitle{Sentiment analysis of twitter data using naive bayes algorithm}.
\bjtitle{International Journal on Recent and Innovation Trends in Computing and
  Communication}
\bvolume{6}(\bissue{4}),
\bfpage{120}--\blpage{125}
(\byear{2018})
\end{barticle}
\endbibitem

\bibitem{Hasan2013}
\begin{barticle}
\bauthor{\bsnm{Hasan}, \binits{M.A.M.}},
\bauthor{\bsnm{Nasser}, \binits{M.}},
\bauthor{\bsnm{Pal}, \binits{B.}}:
\batitle{On the kdd’99 dataset: support vector machine based intrusion
  detection system (ids) with different kernels}.
\bjtitle{Int. J. Electron. Commun. Comput. Eng}
\bvolume{4}(\bissue{4}),
\bfpage{1164}--\blpage{1170}
(\byear{2013})
\end{barticle}
\endbibitem

\bibitem{Hasan2017}
\begin{barticle}
\bauthor{\bsnm{Hasan}, \binits{M.A.M.}},
\bauthor{\bsnm{Ahmad}, \binits{S.}},
\bauthor{\bsnm{Molla}, \binits{M.K.I.}}:
\batitle{Protein subcellular localization prediction using multiple kernel
  learning based support vector machine}.
\bjtitle{Molecular BioSystems}
\bvolume{13}(\bissue{4}),
\bfpage{785}--\blpage{795}
(\byear{2017})
\end{barticle}
\endbibitem

\bibitem{QKuang2009}
\begin{botherref}
\oauthor{\bsnm{Kuang}, \binits{Q.}},
\oauthor{\bsnm{Lei}, \binits{Z.}}:
L.: A practical gpu based knn algorithm.
proceedings of international symposium on computerence \& computational
  technology,
151
(2009)
\end{botherref}
\endbibitem

\bibitem{Kozdrowski2021}
\begin{barticle}
\bauthor{\bsnm{Kozdrowski}, \binits{S.}},
\bauthor{\bsnm{Cichosz}, \binits{P.}},
\bauthor{\bsnm{Paziewski}, \binits{P.}},
\bauthor{\bsnm{Sujecki}, \binits{S.}}:
\batitle{Machine learning algorithms for prediction of the quality of
  transmission in optical networks}.
\bjtitle{Entropy}
\bvolume{23}(\bissue{1}),
\bfpage{7}
(\byear{2021})
\end{barticle}
\endbibitem

\bibitem{Ma2020}
\begin{barticle}
\bauthor{\bsnm{Ma}, \binits{B.}},
\bauthor{\bsnm{Meng}, \binits{F.}},
\bauthor{\bsnm{Yan}, \binits{G.}},
\bauthor{\bsnm{Yan}, \binits{H.}},
\bauthor{\bsnm{Chai}, \binits{B.}},
\bauthor{\bsnm{Song}, \binits{F.}}:
\batitle{Diagnostic classification of cancers using extreme gradient boosting
  algorithm and multi-omics data}.
\bjtitle{Computers in biology and medicine}
\bvolume{121},
\bfpage{103761}
(\byear{2020})
\end{barticle}
\endbibitem

\bibitem{Chang2018}
\begin{barticle}
\bauthor{\bsnm{Chang}, \binits{Y.C.}},
\bauthor{\bsnm{Chang}, \binits{K.H.}},
\bauthor{\bsnm{Wu}, \binits{G.J.}}:
\batitle{Application of extreme gradient boosting trees in the construction of
  credit risk assessment models for financial institutions}.
\bjtitle{Applied Soft Computing}
\bvolume{73},
\bfpage{914}--\blpage{920}
(\byear{2018})
\end{barticle}
\endbibitem

\bibitem{Wang2017}
\begin{barticle}
\bauthor{\bsnm{Wang}, \binits{S.}},
\bauthor{\bsnm{Dong}, \binits{P.}},
\bauthor{\bsnm{Tian}, \binits{Y.}}:
\batitle{A novel method of statistical line loss estimation for distribution
  feeders based on feeder cluster and modified xgboost}.
\bjtitle{Energies}
\bvolume{10}(\bissue{12}),
\bfpage{2067}
(\byear{2017})
\end{barticle}
\endbibitem

\bibitem{Chiu2018}
\begin{barticle}
\bauthor{\bsnm{Chiu}, \binits{D.S.}},
\bauthor{\bsnm{Talhouk}, \binits{A.}}:
\batitle{dicer: an r package for class discovery using an ensemble driven
  approach}.
\bjtitle{BMC bioinformatics}
\bvolume{19},
\bfpage{1}--\blpage{4}
(\byear{2018})
\end{barticle}
\endbibitem

\bibitem{Hand2009}
\begin{barticle}
\bauthor{\bsnm{Hand}, \binits{D.J.}}:
\batitle{Measuring classifier performance: A coherent alternative to the area
  under the roc curve}.
\bjtitle{Machine Learning}
\bvolume{77},
\bfpage{103}--\blpage{123}
(\byear{2009})
\end{barticle}
\endbibitem

\bibitem{Chen2020}
\begin{barticle}
\bauthor{\bsnm{Chen~Q}, \binits{S.R.} \bsuffix{Meng~Z}}
\bjtitle{Frontiers in bioengineering and biotechnology}
\bvolume{8},
\bfpage{1}--\blpage{9}
(\byear{2020})
\end{barticle}
\endbibitem

\bibitem{Salvador-Meneses2019}
\begin{barticle}
\bauthor{\bsnm{Salvador-Meneses}, \binits{J.}},
\bauthor{\bsnm{Ruiz-Chavez}, \binits{Z.}},
\bauthor{\bsnm{Garcia-Rodriguez}, \binits{J.}}:
\batitle{Compressed knn: K-nearest neighbors with data compression}.
\bjtitle{Entropy}
\bvolume{21},
\bfpage{1}--\blpage{20}
(\byear{2019})
\end{barticle}
\endbibitem

\bibitem{Xuekui2020}
\begin{botherref}
\oauthor{\bsnm{Xing}, \binits{L.}},
\oauthor{\bsnm{Joun}, \binits{S.}},
\oauthor{\bsnm{Mackey}, \binits{K.}},
\oauthor{\bsnm{Lesperance}, \binits{M.}},
\oauthor{\bsnm{Zhang}, \binits{X.}}:
Handling high correlations in the feature gene selection using single-cell rna
  sequencing data.
arXiv e-prints,
1--12
(2020)
\end{botherref}
\endbibitem

\end{thebibliography}

\newcommand{\BMCxmlcomment}[1]{}

\BMCxmlcomment{

<refgrp>

<bibl id="B1">
  <title><p>Exponential scaling of single-cell RNA-seq in the past
  decade</p></title>
  <aug>
    <au><snm>Svensson</snm><fnm>V</fnm></au>
    <au><snm>Vento Tormo</snm><fnm>R</fnm></au>
    <au><snm>Teichmann</snm><fnm>S A</fnm></au>
  </aug>
  <source>Nature Protocols</source>
  <pubdate>2018</pubdate>
  <volume>13</volume>
  <fpage>599</fpage>
  <lpage>604</lpage>
</bibl>

<bibl id="B2">
  <title><p>Benchmark and Parameter Sensitivity Analysis of Single-Cell RNA
  Sequencing Clustering Methods</p></title>
  <aug>
    <au><snm>Krzak</snm><fnm>M</fnm></au>
    <au><snm>Raykov</snm><fnm>Y</fnm></au>
    <au><snm>Boukouvalas</snm><fnm>AL</fnm></au>
    <au><snm>Cutillo</snm><fnm>L</fnm></au>
    <au><snm>Claudia</snm><fnm>A</fnm></au>
  </aug>
  <source>Frontiers in Genetics</source>
  <pubdate>2019</pubdate>
  <volume>10</volume>
  <fpage>1</fpage>
  <lpage>19</lpage>
</bibl>

<bibl id="B3">
  <title><p>RNA sequencing: the teenage years</p></title>
  <aug>
    <au><snm>Stark</snm><fnm>R</fnm></au>
    <au><snm>Grzelak</snm><fnm>M</fnm></au>
    <au><snm>Hadfield</snm><fnm>J</fnm></au>
  </aug>
  <source>Nature Reviews Genetics</source>
  <pubdate>2019</pubdate>
  <volume>20</volume>
  <fpage>631</fpage>
  <lpage>656</lpage>
</bibl>

<bibl id="B4">
  <title><p>Overview of Next Generation Sequencing Technologies</p></title>
  <aug>
    <au><snm>Slatko</snm><fnm>B E</fnm></au>
    <au><snm>Gardner</snm><fnm>A F</fnm></au>
    <au><snm>Ausubel</snm><fnm>F M</fnm></au>
  </aug>
  <source>Molecular Biology</source>
  <pubdate>2018</pubdate>
  <volume>122</volume>
  <fpage>1</fpage>
  <lpage>15</lpage>
</bibl>

<bibl id="B5">
  <title><p>A machine learning approach for the identification of key markers
  involved in brain development from single-cell transcriptomic
  data</p></title>
  <aug>
    <au><snm>Hu</snm><fnm>Y</fnm></au>
    <au><snm>Hase</snm><fnm>T</fnm></au>
    <au><snm>Li</snm><fnm>H P</fnm></au>
    <au><snm>Prabhakar</snm><fnm>S</fnm></au>
    <au><snm>Kitano</snm><fnm>H</fnm></au>
    <au><snm>Ng</snm><fnm>S K</fnm></au>
    <au><snm>Ghosh</snm><fnm>S</fnm></au>
    <au><snm>Wee</snm><fnm>L J K</fnm></au>
  </aug>
  <source>BMC genomics</source>
  <pubdate>2016</pubdate>
  <volume>17</volume>
  <fpage>1025</fpage>
</bibl>

<bibl id="B6">
  <title><p>{Method of the Year 2013}</p></title>
  <source>Nature Methods</source>
  <pubdate>2014</pubdate>
  <volume>11</volume>
  <issue>1</issue>
  <fpage>1</fpage>
  <url>https://doi.org/10.1038/nmeth.2801</url>
</bibl>

<bibl id="B7">
  <title><p>Comparative analysis of differential gene expression analysis tools
  for single-cell RNA sequencing data</p></title>
  <aug>
    <au><snm>Wang</snm><fnm>T</fnm></au>
    <au><snm>Li</snm><fnm>B</fnm></au>
    <au><snm>Nelson</snm><fnm>C E</fnm></au>
    <au><snm>Nabavi</snm><fnm>S</fnm></au>
  </aug>
  <source>BMC bioinformatics</source>
  <pubdate>2019</pubdate>
  <volume>20</volume>
  <issue>4</issue>
  <fpage>1</fpage>
  <lpage>16</lpage>
</bibl>

<bibl id="B8">
  <title><p>Identification and Analysis of Glioblastoma Biomarkers Based on
  Single Cell Sequencing</p></title>
  <aug>
    <au><snm>Cheng</snm><fnm>Q</fnm></au>
    <au><snm>Li</snm><fnm>J</fnm></au>
    <au><snm>Fan</snm><fnm>F</fnm></au>
    <au><snm>Cao</snm><fnm>H</fnm></au>
    <au><snm>Dai</snm><fnm>Z Y</fnm></au>
    <au><snm>Wang</snm><fnm>Z Y</fnm></au>
    <au><snm>Feng</snm><fnm>SS</fnm></au>
  </aug>
  <source>Frontiers in Bioengineering and Biotechnology</source>
  <pubdate>2020</pubdate>
  <volume>8</volume>
  <fpage>1</fpage>
  <lpage>7</lpage>
</bibl>

<bibl id="B9">
  <title><p>Clustering and classification methods for single-cell
  RNA-sequencing data</p></title>
  <aug>
    <au><snm>Qi</snm><fnm>R</fnm></au>
    <au><snm>Ma</snm><fnm>A</fnm></au>
    <au><snm>Ma</snm><fnm>Q</fnm></au>
    <au><snm>Zou</snm><fnm>Q</fnm></au>
  </aug>
  <source>Briefings in bioinformatics</source>
  <pubdate>2020</pubdate>
  <volume>21</volume>
  <fpage>1196</fpage>
  <lpage>1208</lpage>
</bibl>

<bibl id="B10">
  <title><p>Supervised classification enables rapid annotation of cell
  atlases</p></title>
  <aug>
    <au><snm>Pliner</snm><fnm>H A</fnm></au>
    <au><snm>Shendure</snm><fnm>J</fnm></au>
    <au><snm>Trapnell</snm><fnm>C</fnm></au>
  </aug>
  <source>Nature Methods</source>
  <pubdate>2019</pubdate>
  <volume>16</volume>
  <fpage>983</fpage>
  <lpage>-986</lpage>
</bibl>

<bibl id="B11">
  <title><p>A comparison of automatic cell identification methods for
  single-cell RNA-sequencing data</p></title>
  <aug>
    <au><snm>Abdelaal</snm><fnm>T</fnm></au>
    <au><snm>Michielsen</snm><fnm>L</fnm></au>
    <au><snm>Cats</snm><fnm>D</fnm></au>
    <au><snm>Hoogduin</snm><fnm>D</fnm></au>
    <au><snm>Mei</snm><fnm>H</fnm></au>
    <au><snm>Reinders</snm><fnm>M</fnm></au>
    <au><snm>Mahfouz</snm><fnm>A</fnm></au>
  </aug>
  <source>Genome Biology</source>
  <pubdate>2019</pubdate>
  <volume>20</volume>
  <fpage>194</fpage>
</bibl>

<bibl id="B12">
  <title><p>Regularization and variable selection via the elastic
  net</p></title>
  <aug>
    <au><snm>Zou</snm><fnm>H</fnm></au>
    <au><snm>Hastie</snm><fnm>T</fnm></au>
  </aug>
  <source>Journal of the royal statistical society: series B (statistical
  methodology)</source>
  <pubdate>2005</pubdate>
  <volume>67</volume>
  <fpage>301</fpage>
  <lpage>320</lpage>
</bibl>

<bibl id="B13">
  <title><p>Linear discriminant analysis</p></title>
  <aug>
    <au><snm>Xanthopoulos</snm><fnm>P</fnm></au>
    <au><snm>Pardalos</snm><fnm>P M</fnm></au>
    <au><snm>Trafalis</snm><fnm>TH B</fnm></au>
  </aug>
  <source>Robust data mining</source>
  <publisher>USA, New York: Springer</publisher>
  <pubdate>2013</pubdate>
  <fpage>27</fpage>
  <lpage>33</lpage>
</bibl>

<bibl id="B14">
  <title><p>Estimating continuous distributions in Bayesian
  classifiers</p></title>
  <aug>
    <au><snm>John</snm><fnm>G H</fnm></au>
    <au><snm>Langley</snm><fnm>P</fnm></au>
  </aug>
  <source>arXiv preprint arXiv:1302.4964</source>
  <pubdate>2013</pubdate>
  <fpage>338</fpage>
  <lpage>345</lpage>
</bibl>

<bibl id="B15">
  <title><p>Support vector machines</p></title>
  <aug>
    <au><snm>Steinwart</snm><fnm>I</fnm></au>
    <au><snm>Christmann</snm><fnm>A</fnm></au>
  </aug>
  <publisher>Germany: Springer Science \& Business Media</publisher>
  <pubdate>2008</pubdate>
</bibl>

<bibl id="B16">
  <title><p>K-nearest neighbors</p></title>
  <aug>
    <au><snm>Kramer</snm><fnm>O</fnm></au>
  </aug>
  <source>Dimensionality reduction with unsupervised nearest neighbors</source>
  <publisher>USA, New York: Springer</publisher>
  <pubdate>2013</pubdate>
  <fpage>13</fpage>
  <lpage>23</lpage>
</bibl>

<bibl id="B17">
  <title><p>Performance analysis of classification tree learning
  algorithms</p></title>
  <aug>
    <au><snm>Gupta</snm><fnm>DL</fnm></au>
    <au><snm>Malviya</snm><fnm>AK</fnm></au>
    <au><snm>Singh</snm><fnm>S</fnm></au>
  </aug>
  <source>International Journal of Computer Applications</source>
  <pubdate>2012</pubdate>
  <volume>55</volume>
  <issue>6</issue>
  <fpage>39</fpage>
  <lpage>44</lpage>
</bibl>

<bibl id="B18">
  <title><p>Multiple additive regression trees with application in
  epidemiology</p></title>
  <aug>
    <au><snm>Friedman</snm><fnm>J H</fnm></au>
    <au><snm>Meulman</snm><fnm>J J</fnm></au>
  </aug>
  <source>Statistics in medicine</source>
  <pubdate>2003</pubdate>
  <volume>22</volume>
  <fpage>1365</fpage>
  <lpage>1381</lpage>
</bibl>

<bibl id="B19">
  <title><p>Ensemble learning</p></title>
  <aug>
    <au><snm>Dietterich</snm><fnm>T G</fnm></au>
  </aug>
  <source>The handbook of brain theory and neural networks</source>
  <pubdate>2002</pubdate>
  <volume>2</volume>
  <fpage>110</fpage>
  <lpage>125</lpage>
</bibl>

<bibl id="B20">
  <title><p>Bias, robustness and scalability in single-cell differential
  expression analysis</p></title>
  <aug>
    <au><snm>Soneson</snm><fnm>C</fnm></au>
    <au><snm>Robinson</snm><fnm>M D</fnm></au>
  </aug>
  <source>Nature Methods</source>
  <pubdate>2018</pubdate>
  <volume>15</volume>
  <fpage>255</fpage>
  <lpage>261</lpage>
</bibl>

<bibl id="B21">
  <title><p>A lineage-resolved molecular atlas of C. Elegans embryogenesis at
  single-cell resolution</p></title>
  <aug>
    <au><snm>Packer</snm><fnm>J S</fnm></au>
    <au><snm>Zhu</snm><fnm>Q</fnm></au>
    <au><snm>Huynh</snm><fnm>C</fnm></au>
    <au><snm>Sivaramakrishnan</snm><fnm>P</fnm></au>
    <au><snm>Preston</snm><fnm>E</fnm></au>
    <au><snm>Dueck</snm><fnm>H</fnm></au>
    <au><snm>Stefanik</snm><fnm>D</fnm></au>
    <au><snm>Tan</snm><fnm>K</fnm></au>
    <au><snm>Trapnell</snm><fnm>C</fnm></au>
    <au><snm>Kim</snm><fnm>J</fnm></au>
    <au><snm>Waterston</snm><fnm>R H</fnm></au>
    <au><snm>Murray</snm><fnm>J I</fnm></au>
  </aug>
  <source>Science</source>
  <pubdate>2019</pubdate>
  <volume>365</volume>
  <fpage>1</fpage>
  <lpage>15</lpage>
</bibl>

<bibl id="B22">
  <aug>
    <au><snm>Soneson</snm><fnm>C</fnm></au>
    <au><snm>Robinson</snm><fnm>MD</fnm></au>
  </aug>
  <source>\url{http://imlspenticton.uzh.ch/robinson_lab/conquer_de_comparison}</source>
</bibl>

<bibl id="B23">
  <title><p>PowsimR: Power analysis for bulk and single cell RNA-seq
  experiments</p></title>
  <aug>
    <au><snm>Vieth</snm><fnm>B</fnm></au>
    <au><snm>Ziegenhain</snm><fnm>C</fnm></au>
    <au><snm>Parekh</snm><fnm>S</fnm></au>
    <au><snm>Enard</snm><fnm>W</fnm></au>
    <au><snm>Hellmann</snm><fnm>I</fnm></au>
  </aug>
  <source>Bioinformatics</source>
  <pubdate>2017</pubdate>
  <volume>33</volume>
  <issue>21</issue>
  <fpage>3486–3488</fpage>
</bibl>

<bibl id="B24">
  <title><p>Bilayer elastic net regression model for supervised
  spectral-spatial hyperspectral image classification</p></title>
  <aug>
    <au><snm>Soomro</snm><fnm>B N</fnm></au>
    <au><snm>Xiao</snm><fnm>L</fnm></au>
    <au><snm>Huang</snm><fnm>L</fnm></au>
    <au><snm>Soomro</snm><fnm>SH</fnm></au>
    <au><snm>Molaei</snm><fnm>M</fnm></au>
  </aug>
  <source>IEEE Journal of Selected Topics in Applied Earth Observations and
  Remote Sensing</source>
  <pubdate>2016</pubdate>
  <volume>9</volume>
  <issue>9</issue>
  <fpage>4102</fpage>
  <lpage>4116</lpage>
</bibl>

<bibl id="B25">
  <title><p>Drug-induced cell viability prediction from LINCS-L1000 through
  WRFEN-XGBoost algorithm</p></title>
  <aug>
    <au><snm>Lu</snm><fnm>J</fnm></au>
    <au><snm>Chen</snm><fnm>M</fnm></au>
    <au><snm>Qin</snm><fnm>Y</fnm></au>
  </aug>
  <source>BMC bioinformatics</source>
  <pubdate>2021</pubdate>
  <volume>22</volume>
  <issue>1</issue>
  <fpage>1</fpage>
  <lpage>18</lpage>
</bibl>

<bibl id="B26">
  <title><p>A comparison of generalized linear discriminant analysis
  algorithms</p></title>
  <aug>
    <au><snm>Park</snm><fnm>C H</fnm></au>
    <au><snm>Park</snm><fnm>H</fnm></au>
  </aug>
  <source>Pattern Recognition</source>
  <pubdate>2008</pubdate>
  <volume>41</volume>
  <issue>3</issue>
  <fpage>1083</fpage>
  <lpage>1097</lpage>
</bibl>

<bibl id="B27">
  <title><p>Linear discriminant analysis: A detailed tutorial</p></title>
  <aug>
    <au><snm>Tharwat</snm><fnm>A</fnm></au>
    <au><snm>Gaber</snm><fnm>T</fnm></au>
    <au><snm>Ibrahi</snm><fnm>A</fnm></au>
    <au><snm>Hassanien</snm><fnm>A E</fnm></au>
  </aug>
  <source>AI communications</source>
  <pubdate>2017</pubdate>
  <volume>30</volume>
  <issue>2</issue>
  <fpage>169</fpage>
  <lpage>190</lpage>
</bibl>

<bibl id="B28">
  <title><p>Sentiment Analysis of Twitter Data Using Naive Bayes
  Algorithm</p></title>
  <aug>
    <au><snm>Malik</snm><fnm>V</fnm></au>
    <au><snm>Kumar</snm><fnm>A</fnm></au>
  </aug>
  <source>International Journal on Recent and Innovation Trends in Computing
  and Communication</source>
  <pubdate>2018</pubdate>
  <volume>6</volume>
  <issue>4</issue>
  <fpage>120</fpage>
  <lpage>125</lpage>
</bibl>

<bibl id="B29">
  <title><p>On the KDD’99 dataset: support vector machine based intrusion
  detection system (ids) with different kernels</p></title>
  <aug>
    <au><snm>Hasan</snm><fnm>MD AL M</fnm></au>
    <au><snm>Nasser</snm><fnm>M</fnm></au>
    <au><snm>Pal</snm><fnm>B</fnm></au>
  </aug>
  <source>Int. J. Electron. Commun. Comput. Eng</source>
  <pubdate>2013</pubdate>
  <volume>4</volume>
  <issue>4</issue>
  <fpage>1164</fpage>
  <lpage>1170</lpage>
</bibl>

<bibl id="B30">
  <title><p>Protein subcellular localization prediction using multiple kernel
  learning based support vector machine</p></title>
  <aug>
    <au><snm>Hasan</snm><fnm>MD AL M</fnm></au>
    <au><snm>Ahmad</snm><fnm>SH</fnm></au>
    <au><snm>Molla</snm><fnm>MD K I</fnm></au>
  </aug>
  <source>Molecular BioSystems</source>
  <pubdate>2017</pubdate>
  <volume>13</volume>
  <issue>4</issue>
  <fpage>785</fpage>
  <lpage>795</lpage>
</bibl>

<bibl id="B31">
  <title><p>L.: A practical GPU based KNN algorithm</p></title>
  <aug>
    <au><snm>Kuang</snm><fnm>Q.</fnm></au>
    <au><snm>Lei</snm><fnm>Z.</fnm></au>
  </aug>
  <source>proceedings of international symposium on computerence \&
  computational technology</source>
  <pubdate>2009</pubdate>
  <fpage>151</fpage>
</bibl>

<bibl id="B32">
  <title><p>Machine Learning Algorithms for Prediction of the Quality of
  Transmission in Optical Networks</p></title>
  <aug>
    <au><snm>Kozdrowski</snm><fnm>S</fnm></au>
    <au><snm>Cichosz</snm><fnm>P</fnm></au>
    <au><snm>Paziewski</snm><fnm>P</fnm></au>
    <au><snm>Sujecki</snm><fnm>S</fnm></au>
  </aug>
  <source>Entropy</source>
  <pubdate>2021</pubdate>
  <volume>23</volume>
  <issue>1</issue>
  <fpage>7</fpage>
</bibl>

<bibl id="B33">
  <title><p>Diagnostic classification of cancers using extreme gradient
  boosting algorithm and multi-omics data</p></title>
  <aug>
    <au><snm>Ma</snm><fnm>B</fnm></au>
    <au><snm>Meng</snm><fnm>F</fnm></au>
    <au><snm>Yan</snm><fnm>G</fnm></au>
    <au><snm>Yan</snm><fnm>H</fnm></au>
    <au><snm>Chai</snm><fnm>B</fnm></au>
    <au><snm>Song</snm><fnm>F</fnm></au>
  </aug>
  <source>Computers in biology and medicine</source>
  <pubdate>2020</pubdate>
  <volume>121</volume>
  <fpage>103761</fpage>
</bibl>

<bibl id="B34">
  <title><p>Application of eXtreme gradient boosting trees in the construction
  of credit risk assessment models for financial institutions</p></title>
  <aug>
    <au><snm>Chang</snm><fnm>YC</fnm></au>
    <au><snm>Chang</snm><fnm>K H</fnm></au>
    <au><snm>Wu</snm><fnm>G J</fnm></au>
  </aug>
  <source>Applied Soft Computing</source>
  <pubdate>2018</pubdate>
  <volume>73</volume>
  <fpage>914</fpage>
  <lpage>920</lpage>
</bibl>

<bibl id="B35">
  <title><p>A novel method of statistical line loss estimation for distribution
  feeders based on feeder cluster and modified XGBoost</p></title>
  <aug>
    <au><snm>Wang</snm><fnm>S</fnm></au>
    <au><snm>Dong</snm><fnm>P</fnm></au>
    <au><snm>Tian</snm><fnm>Y</fnm></au>
  </aug>
  <source>Energies</source>
  <pubdate>2017</pubdate>
  <volume>10</volume>
  <issue>12</issue>
  <fpage>2067</fpage>
</bibl>

<bibl id="B36">
  <title><p>diceR: an R package for class discovery using an ensemble driven
  approach</p></title>
  <aug>
    <au><snm>Chiu</snm><fnm>D S</fnm></au>
    <au><snm>Talhouk</snm><fnm>A</fnm></au>
  </aug>
  <source>BMC bioinformatics</source>
  <pubdate>2018</pubdate>
  <volume>19</volume>
  <fpage>1</fpage>
  <lpage>4</lpage>
</bibl>

<bibl id="B37">
  <title><p>Measuring classifier performance: A coherent alternative to the
  area under the ROC curve</p></title>
  <aug>
    <au><snm>Hand</snm><fnm>D J</fnm></au>
  </aug>
  <source>Machine Learning</source>
  <pubdate>2009</pubdate>
  <volume>77</volume>
  <fpage>103</fpage>
  <lpage>123</lpage>
</bibl>

<bibl id="B38">
  <aug>
    <au><snm>Chen Q</snm><fnm>SR</fnm></au>
  </aug>
  <source>Frontiers in bioengineering and biotechnology</source>
  <pubdate>2020</pubdate>
  <volume>8</volume>
  <fpage>1–9</fpage>
</bibl>

<bibl id="B39">
  <title><p>Compressed kNN: K-nearest neighbors with data
  compression</p></title>
  <aug>
    <au><snm>Salvador Meneses</snm><fnm>J</fnm></au>
    <au><snm>Ruiz Chavez</snm><fnm>Z</fnm></au>
    <au><snm>Garcia Rodriguez</snm><fnm>J</fnm></au>
  </aug>
  <source>Entropy</source>
  <pubdate>2019</pubdate>
  <volume>21</volume>
  <fpage>1</fpage>
  <lpage>20</lpage>
</bibl>

<bibl id="B40">
  <title><p>Handling high correlations in the feature gene selection using
  Single-Cell RNA sequencing data</p></title>
  <aug>
    <au><snm>Xing</snm><fnm>L</fnm></au>
    <au><snm>Joun</snm><fnm>S</fnm></au>
    <au><snm>Mackey</snm><fnm>K</fnm></au>
    <au><snm>Lesperance</snm><fnm>M</fnm></au>
    <au><snm>Zhang</snm><fnm>X</fnm></au>
  </aug>
  <source>arXiv e-prints</source>
  <pubdate>2020</pubdate>
  <fpage>1</fpage>
  <lpage>12</lpage>
</bibl>

</refgrp>
} 




\section*{Figures}
  \begin{figure}[!ht]
   \includegraphics[width=11.2cm]{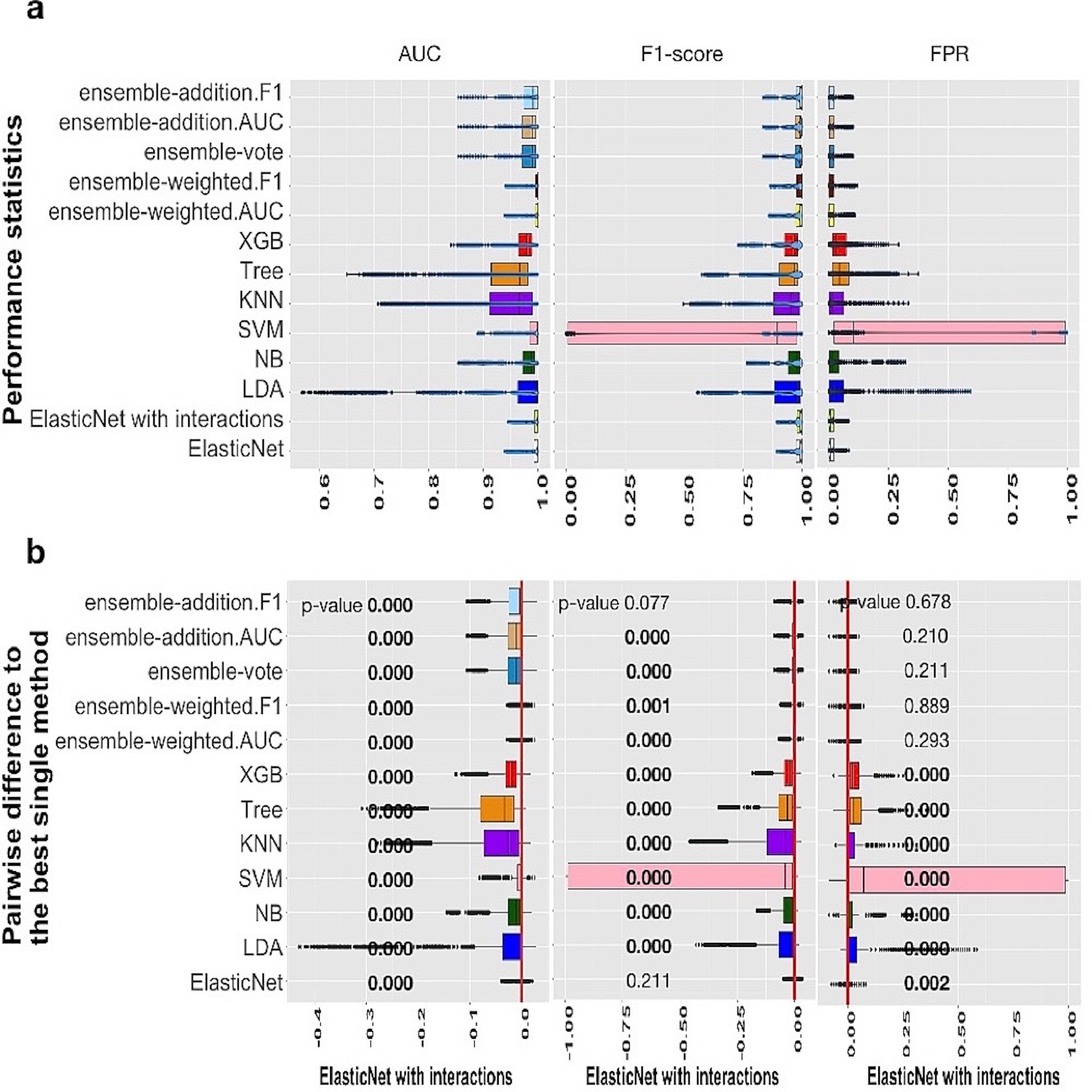}
  \caption{Performance of 13 algorithms on 27 small data sets. The values of the three criteria (AUC, F1-score, and FPR) are expressed on 27 small data sets in \textbf{a}, which shows AUC, F1-score, and FPR, from left to right. There are 2700 points in each box. The values of three criteria were used to represent the performance of the corresponding algorithm. \textbf{b} represents the difference between the other algorithms and the best single algorithm. From left to right, the results display the difference under AUC, F1, and FPR. We added the p-values of the Wilcoxon test in \textbf{b}, the p-value over $<$ 0.05 is bolder, and the red line presents the zero value. The best single method is shown at the bottom of the boxes.}
  \label{fig1}
\end{figure}

\begin{figure}[!ht]
\includegraphics[width=11.2cm]{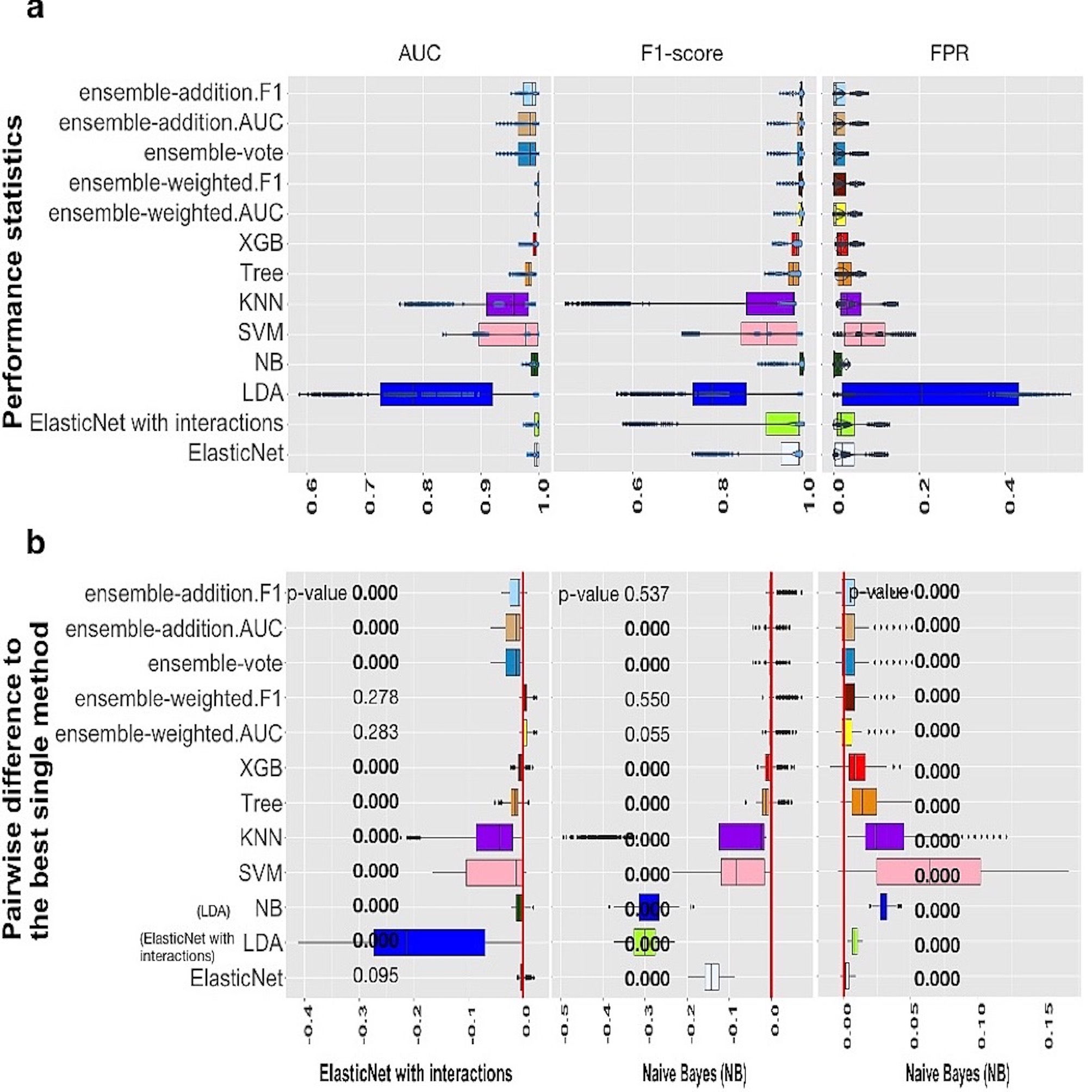}
  \caption{Performance of 13 algorithms on 4 medium data sets. The values of the three criteria (AUC, F1-score, and FPR) are expressed on 4 medium data sets in \textbf{a}, which shows AUC, F1-score, and FPR, from left to right. There are 400 points in each box. The values of three criteria were used to represent the performance of the corresponding algorithm. \textbf{b} represents the difference between the other algorithms and the best single algorithm. From left to right, the results display the difference under AUC, F1-score, and FPR. We added the p-values of the Wilcoxon test in \textbf{b}, the p-value over $<$ 0.05 is bolder, and the red line presents the zero value. The best single methods are shown at the bottom of the boxes. Because the two methods were the best single ones, we have marked the algorithm names under the F1 and FPR  standards in the coordinate axis brackets.}
\label{fig2}
\end{figure}

\begin{figure}[htb]
\includegraphics[width=11.2cm]{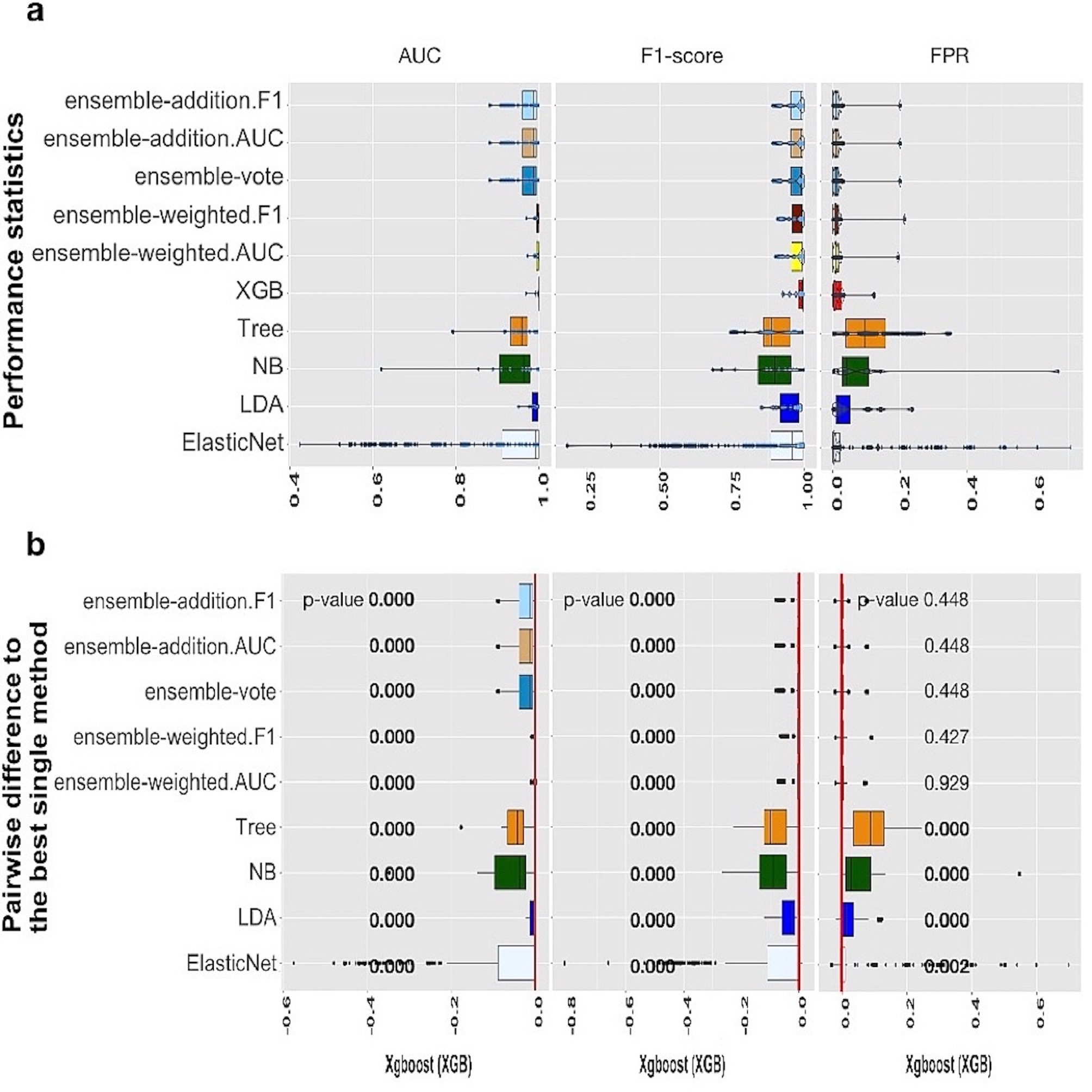}
  \caption{Performance of 10 algorithms on 12 large data sets. The values of the three criteria (AUC, F1-score, and FPR) are expressed on 12 large data sets in \textbf{a}, which shows AUC, F1-score, and FPR, from left to right. There are 1200 points in each box. The values of three criteria were used to represent the performance of the corresponding algorithm. \textbf{b} represents the difference between the other algorithms and the best single algorithm. From left to right, the results display the difference under AUC, F1, and FPR. We added the p-values of the Wilcoxon test in \textbf{b}, the p-value over $<$ 0.05 is bolder, and the red line presents the zero value. The best single method is shown at the bottom of the boxes.}
  \label{fig3}
\end{figure}

  \begin{figure}[!h]
  \includegraphics[width=11.2cm]{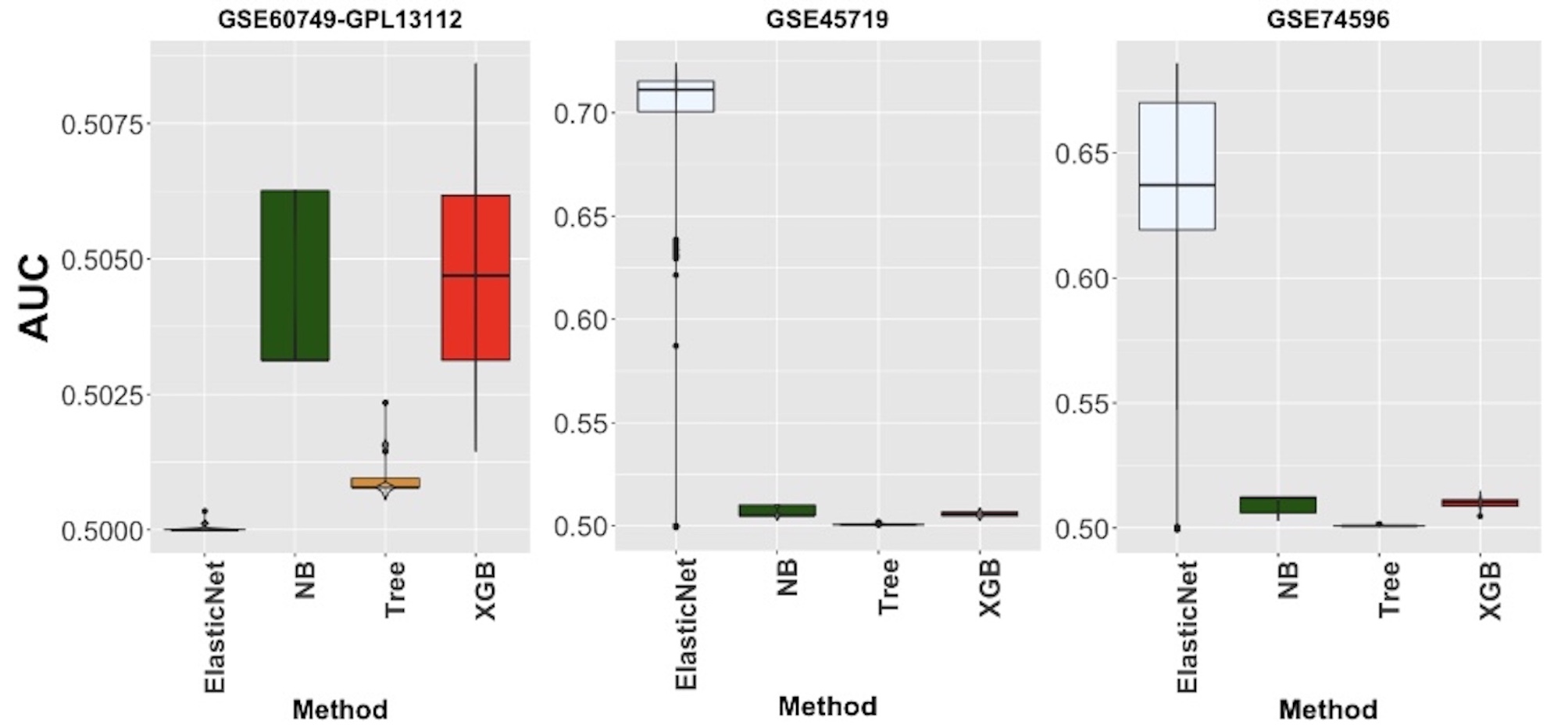}
  \caption{Gene selection of 4 algorithms on 100 subsets of each simulated data set. There are three data sets from left to right, and we have marked the data set names at the top. Each box shows the AUC values of the 4 algorithms after 100-times gene selections on the simulated data set. The horizontal axis is arranged by the algorithms name.}
  \label{fig4}
\end{figure}

\vspace{5cm}
\section*{Tables}
\begin{table}[!ht]\tiny
    \caption{The computation time of an algorithm on a specific data set. Each column represents a different algorithm, and the rows indicate different data sets. We annotated the sample size of the data sets in the table. The maximum speed values were illustrated by underline.}\label{Table5}
    \begin{tabular}{llllllllll}
    \hline
         {\textbf{Label}}& \textbf{ElasticNet} &\begin{tabular}[c]{@{}l@{}}
\textbf{ElasticNet}\\ \textbf{with}\\\textbf{interactions}
\end{tabular} & \textbf{LDA} & \textbf{NB} & \textbf{SVM} & \textbf{KNN} & \textbf{Tree} & \textbf{XGB} &\begin{tabular}[c]{@{}l@{}}
\textbf{Sample}\\ \textbf{size}
\end{tabular}\\ \hline
       \multicolumn{9}{c}{\textbf{The computation time in 27 small data sets (seconds)}} \\
       \hline
         1 & 37.057&60.918 & \underline{1.969}& 7.136 &639.367 & 9.681 & 15.915 & 29.133 & 192\\ 
        2 & 14.362& 24.385& \underline{0.980} & 5.307& 247.713 & 3.990 & 8.028 & 12.114 &95 \\ 
        3 & 198.204 &30.126& \underline{5.943} & 14.138 &5208.412 & 68.672 & 35.271 & 107.450 &564\\ 
        4 & 24.998& 38.792 &\underline{0.996} & 5.723 & 290.221 & 4.500 & 9.270 & 13.388 &110 \\ 
        5 & 40.536 &  65.178&\underline{1.682} & 7.300 & 632.190 & 9.118 & 11.990 & 22.355 &192 \\ 
        6 & 38.661 & 64.686& \underline{1.626} & 7.203 & 615.368 & 8.911 & 10.710 & 19.079&186 \\ 
        7 & 20.004 &29.290&\underline{0.877} & 5.219 & 240.178 & 3.889 & 9.563 & 13.565 &99\\ 
        8 & 40.087 &61.180& \underline{1.661} & 6.916 & 650.740 & 9.053 & 14.454 & 26.990 &192\\ 
        9 & 37.518 &35.188& \underline{1.494} & 6.675 & 411.244 & 7.798 & 10.358 & 18.767 &91 \\ 
        10 & 80.411 &121.676& \underline{2.319} & 8.260 & 835.417 & 15.678 & 12.739 & 25.214&268 \\
        11 & 37.893 &58.476 & \underline{1.283}& 6.141 & 339.512 & 6.335 & 10.064 & 16.080 &147\\ 
        12 & 430.164 &170.021& \underline{9.560}& 18.190 & 10328.148 & 161.913 & 18.037 & 121.704 &192\\ 
        13 & 104.301 & 111.083& \underline{1.631}& 7.095 &673.008 & 8.861 & 10.582 & 34.456 &192\\ 
        14 & 100.775 &111.154&\underline{2.659}& 9.619 & 1545.478 & 22.382 & 14.804 & 37.964 &328 \\ 
        15 & 116.645 &125.838 &\underline{3.809}& 11.827 & 3184.677 & 46.696 & 26.341 & 58.991 &460 \\ 
        16 & 56.745 &65.334& \underline{1.440}& 7.215 &553.960 & 9.138& 11.650 & 21.656&183 \\
        17 & 353.498 &33.495& \underline{7.221}& 16.828 & 1288.900 & 126.284 & 34.363 & 90.167 &106 \\ 
        18 & 96.279 &101.553&\underline{2.455}& 9.689 & 1539.189 & 21.260 & 19.310 & 37.981&313 \\ 
        19 & 35.485 &25.004&\underline{0.897}& 6.041 & 277.117 & 4.730 & 9.726 & 15.810&91 \\
        20 & 25.816&27.386 &\underline{0.760}& 5.652 & 238.028 & 4.020 & 7.990 & 12.259&96 \\ 
        21 & 178.537 &164.780&\underline{1.456} & 7.561 & 440.966 & 9.325 & 8.794 & 19.780&188 \\ 
        22 & 43.940 &50.828&\underline{1.403}& 7.359 & 566.504 & 9.146 & 10.180 & 18.819&181 \\ 
        23 & 72.664 &83.030&\underline{1.502}& 7.613 & 657.644 & 9.504 & 9.713 & 27.458 &192 \\ 
        24 & 66.112 &96.200& \underline{1.487}& 7.616 & 639.454 & 9.488 & 10.289 & 27.005 & 192\\ 
        25 & 59.439 &86.527& \underline{1.528}& 7.638 & 664.293 & 9.681 & 13.983 & 29.074 &192\\
        26 & 126.590 &85.750&\underline{5.288}& 14.354 & 3649.444 & 74.889 & 33.485 & 78.146&269 \\ 
        27 & 30.640 & 53.641&\underline{1.339}& 7.032 & 481.385 & 7.598 & 12.044 & 20.131 &164\\ \hline
         \multicolumn{9}{c}{\textbf{The computation time in 4 medium data sets (seconds)}}  \\ \hline
        1 & 163.504 & 236.97 & \underline{11.765} & 23.030 & 4194.281 & 176.300 & 32.935 & 90.274 & 843 \\ 
        2 & 233.984 & 670.67 & \underline{16.954} & 29.639 & 9350.588 & 325.453 & 55.913 & 139.470 & 1126\\ 
        3 & 640.719 & 46.010  & \underline{14.902} & 22.276 & 12406.980 & 263.876 & 48.200 & 121.668 & 992  \\ 
        4 & 41.442 & 164.89 & \underline{2.660} & 10.531 & 518.837 & 22.619& 14.459 & 27.354 &  234\\ \hline
         \multicolumn{9}{c}{\textbf{The computation time in 12 large data sets (seconds)}} \\ \hline
        1 &  485.537 & NA & \underline{247.775} & 391.010 & NA & NA & 724.572 & 2566.135 & 19252 \\ 
        2 & 258.447 &NA & \underline{249.383} & 400.039 & NA & NA & 676.196 & 1792.239 &  5328 \\
        3 & 2276.346&NA & \underline{282.598} & 380.155 & NA & NA & 920.946 & 1910.279 & 21492 \\ 
        4 & 1932.343 &NA & \underline{235.466} & 405.532 & NA & NA & 626.151 & 2986.626  & 22661  \\ 
        5 & 9093.946 &NA & \underline{165.728} & 267.974 & NA & NA & 357.760 & 1150.460 & 23718 \\ 
        6 & 9026.280 &NA & \underline{510.495} & 741.520 & NA & NA & 1469.216 & 7039.001 & 20213 \\ 
        7 & 31978.390 &NA & \underline{370.260} & 492.451 & NA & NA & 913.547 & 3274.896 & 13577 \\ 
        8 & 7256.621 &NA & \underline{346.646} & 520.316 & NA & NA & 923.122 & 2940.168 & 33043 \\ 
        9 &10807.530 &NA & \underline{466.657} & 667.569 & NA & NA & 1163.817 & 3267.698 & 27700 \\ 
        10 & 1961.241 & NA& \underline{384.110} & 551.439 & NA & NA & 1242.170 & 3313.920 & 28757 \\ 
        11 & 11386.465 & NA& \underline{233.273} & 434.554 & NA & NA & 829.204 & 2607.679 & 36407 \\ 
        12 & 2691.988 & NA& \underline{65.735} & 116.332 & NA & NA & 231.935 & 792.720 & 37464 \\ \hline
    \end{tabular}
\end{table} 


\section*{Supplementary Information}
\vspace{0.2cm}
  \subsection*{Additional file 1 --- Figure S1, Figure S2 and Figure S3. }
  Figure S1.The values of each criterion, Median values, and IQR values of the AUC, F1-score, FPR, Recall, and Precision are indicated for 27 small data sets. Figure S2. The values of each criterion, Median values, and IQR values of the AUC, F1-score, FPR, Recall, and Precision are indicated for 4 medium data sets. Figure S3. The values of each criterion, Median values, and IQR values of the AUC, F1-score, FPR, Recall, and Precision are indicated for 12 large data sets. 
  \subsection*{Additional file 2 --- Table S1.}
  27 data sets from conquer
 \subsection*{Additional file 3 --- Table S2.}
 4 filtered subsets from Packer et al.
 \subsection*{Additional file 4 --- Table S3.}
12 filtered subsets from Packer et al.
 \subsection*{Additional file 5 --- Table S4.}
3 simulated data sets provided by Soneson et al.

\end{backmatter}
\end{document}